\documentclass{aa} 
\usepackage{graphicx}
\usepackage[varg]{txfonts}
\usepackage{natbib}
\usepackage{txfonts}
\usepackage[dvipsnames]{xcolor}
\usepackage[colorlinks=true, linkcolor={blue!70!black}, citecolor={blue!70!black}, urlcolor={blue!70!black}]{hyperref}

\usepackage{amssymb}
\usepackage{pifont}
\newcommand{\cmark}{\ding{51}}%
\newcommand{\xmark}{\ding{55}}%

\newcommand{\kms}{km s$^{-1}$}

\newcommand{\bbarolo}{\textsc{$^{\rm 3D}$Barolo}}

\usepackage{CJKutf8}
\begin{document} 

   \title{ \textit{JWST} + ALMA ubiquitously discover companion systems within $\lesssim18\,$kpc around four $z$$\approx$3.5 luminous radio-loud AGN }
   \titlerunning{\textit{JWST} + ALMA discover nearby companions around $z$$\approx$3.5 radio AGN}

\author{Wuji Wang (\begin{CJK*}{UTF8}{gbsn}王无忌\end{CJK*})   \inst{\ref{instHD},\ref{instCal}}
        \and 
        Carlos De Breuck\inst{\ref{instESO}}
        \and
        Dominika Wylezalek\inst{\ref{instHD}}
        \and
        Jo\"{e}l Vernet\inst{\ref{instESO}}
        \and 
        Matthew D. Lehnert\inst{\ref{insML}}
        \and
        Daniel Stern\inst{\ref{instDS}}
        \and
        David S. N. Rupke\inst{\ref{instDR},\ref{instHD}}
        \and
        Nicole P. H. Nesvadba\inst{\ref{instNN}}
        \and
        Andrey Vayner\inst{\ref{instCal}}
        \and
        Nadia L. Zakamska\inst{\ref{instJHU}}  
        \and
        Lingrui Lin (\begin{CJK*}{UTF8}{gbsn}林令瑞\end{CJK*}) \inst{\ref{instNJU1},\ref{instNJU2}}
        \and
        Pranav Kukreti\inst{\ref{instHD}}
        \and
        Bruno Dall'Agnol de Oliveira\inst{\ref{instHD}}
        \and
        Julian T. Groth\inst{\ref{instHD}}   
        }

    \institute{Zentrum f\"{u}r Astronomie der Universit\"{a}t Heidelberg, Astronomisches Rechen-Institut, M\"{o}nchhofstr. 12-14, D-69120 Heidelberg, Germany\label{instHD}\\
                \email{wujiwang@ipac.caltech.edu}$;\,\,$\email{wuji.wang\_astro@outlook.com}
        \and
            IPAC, M/C 314-6, California Institute of Technology, 1200 East California Boulevard, Pasadena, CA 91125, USA\label{instCal}
   \and
   European Southern Observatory, Karl-Schwarzchild-Str. 2, D-85748 Garching, Germany\label{instESO}
   \and 
    Univ. Lyon, Univ. Lyon1, ENS de Lyon, CNRS, Centre de Recherche Astrophysique de Lyon UMR5574, 69230 Saint-Genis-Laval, France\label{insML}
    \and
   Jet Propulsion Laboratory, California Institute of Technology, 4800 Oak Grove
Drive, Pasadena, CA91109, USA\label{instDS}
    \and
    Department of Physics, Rhodes College, 2000 N. Parkway, Memphis, TN 38104, USA\label{instDR}
   \and
   Universit\'{e} de la C\^{o}te d’Azur, Observatoire de la C\^{o}te d’Azur, CNRS, Laboratoire Lagrange, Bd de l’Observatoire, CS 34229, F-06304 Nice cedex 4, France\label{instNN}
    \and
   Department of Physics and Astronomy, Bloomberg Center, Johns Hopkins University, 3400 N. Charles Street, Baltimore, MD 21218, USA\label{instJHU}
    \and
   School of Astronomy and Space Science, Nanjing University, Nanjing, 210000, China\label{instNJU1}
   \and
   Key Laboratory of Modern Astronomy and Astrophysics, Nanjing University, Ministry of Education, Nanjing 210093, China\label{instNJU2}
        }
   \date{Received 3rd January 2025 / Accepted 26th February 2025}
\authorrunning{Wuji Wang et al.}
 
  \abstract{Mergers play important roles in galaxy evolution at and beyond Cosmic Noon ($z\sim3$). They are found to be a trigger of active galactic nuclei (AGN) activity and a process for growing stellar mass and black hole mass. High-$z$ radio galaxies (HzRGs=type-2 radio-loud AGN) are among the most massive galaxies known, and reside in dense environments on scales of tens of kiloparsecs to Megaparsecs. We present the first search for kpc-scale companions in a sample of four $z\sim3.5$ HzRGs, with many supporting datasets, using matched 0.2\arcsec resolution ALMA and \textit{JWST}/NIRSpec integral field unit data.  
  We discover a total of $\sim12$ companion systems within $\lesssim18\,$kpc across all four HzRG fields using two independent detection methods: peculiar [\ion{O}{iii}]$4959,5007$ kinematics offset from the main (systemic) ionized gas component and [\ion{C}{ii}]$158\rm \mu m$ emitters. We examine the velocity fields of these companions and find evidence of disk rotation along with more complex motions. We estimate the dynamical masses of these nearby systems to be $M_{\rm dyn}\sim10^{9-11}\,M_{\odot}$, which may indicate a minor merger scenario. Our results indicate that these companions may be the trigger of the powerful radio-loud AGN. We discuss the roles of the discovered companion systems in galaxy evolution for these powerful jetted AGN and indicate that they may impede jet launch and deflect the jet.
  } 

   \keywords{   Galaxies: evolution --
                galaxies: high-redshift --
                galaxies: jets --
                galaxies: ISM --
                quasars: emission lines
               }

   \maketitle

%

\section{Introduction}\label{sec:introdc}
The hierarchical model for galaxy evolution suggests that galaxy-galaxy mergers are one of the important mechanisms in stellar mass build-up \citep[e.g.,][]{Moster_2018,Behroozi_2019}. The fraction of mergers, $f_{\mathrm{MM}}$, is found to be higher in the early universe \citep[e.g., $f_{\mathrm{MM}}\sim0.44$ at $z \sim 4.5$ through large ALMA survey targeting star forming galaxies,][]{Romano_2021}. The merger fraction evolution mirrors the trend of star formation rate (SFR) density through cosmic history with a peak around $z\sim3$ \citep[i.e., Cosmic Noon,][]{MandD_2014,Husko_2022}. Given that the growth rate of supermassive black holes (SMBHs) related to active galactic nuclei (AGN) activity, also follows the cosmic SFR density evolution \citep[][]{Heckman_2014}, it is reasonable to conclude that the galaxy-galaxy interactions play an important role in feeding black holes and building stellar mass by either direct collapse or stimulating star formation. Indeed, simulations show that the merger processes enhance starbursts \citep[][]{Hopkins_2008d,Hopkins_2008}. Quasar activity is also shown to be related to mergers, suggesting that they play an important role in both the feeding and feedback processes \citep[e.g.,][]{Pierce_2023}.

Early works studying mergers were hampered by limited resolution and sensitivity. The detection of mergers is difficult without kinematic information. The successful operation of \textit{JWST} provided the opportunity for a clearer look. Using NIRCam, \citet{Gupta_2023} reported that the extreme emission lines observed at $z\sim3$ could be driven by major mergers. \citet{Dalmasso_2024} quantified the morphologies of $z>4$ galaxies and reported the fraction rate of mergers even beyond the epoch of reionization. In addition to imaging, NIRSpec is a more powerful tool for merger studies especially with its integral field unit (IFU) mode. For example, \citet{Jones_2024} studied the ionized gas kinematics of a dense cluster system at $z\sim6.34$ with the NIRSpec IFU. Resolution-matched observations by Atacama Large Millimeter/submillimeter Array (ALMA) are equally important for high-redshift companion studies. For example, the far-infrared (FIR) line [\ion{C}{II}]$158\rm \mu m$ (hereinafter [\ion{C}{ii}]), redshifted into the spectral window of ALMA, is a more reliable tracer of cold gas dynamics than the warm ionized optical lines which are affected by strong AGN \citep[e.g.,][]{Lagache_2018,Dessauges-Zavadsky_2020}. Simultaneously, the continuum underneath offers us information on the dust content of the galaxies whose detections and morphologies are very useful in the context of merger studies \citep[e.g.,][]{Faisst_2020,Romano_2021}.

It is of significant interest to look at the so called transition period from Cosmic Dawn to Noon (i.e., $3\lesssim z \lesssim 6$) and study how the growth of SMBHs and star formation are reaching their peak. This period of time is important to look for the quenching of some of the most massive galaxies and energetic AGN \citep[e.g.,][]{Suzuki_2022}. Among the quasar species, high-redshift radio galaxies \citep[HzRGs,][]{Miley_2008} uniquely useful for investigating close companions and mergers. HzRGs are type-2 quasars, thereby offering a clean view of the vicinity around the host galaxy with the ``natural coronograph" blocking the bright quasar \citep[e.g.,][]{vernet_2001}. It has been proposed that mergers are a possible trigger of the powerful jet (radio-loud AGN), but the link is elusive \citep[][]{Chiaberge_2015}. With joint \textit{JWST} and ALMA observations, we can search for companions, and merger candidates, on tens of kiloparsec scales, moving a step towards understanding the ignition of HzRGs and their impact on galaxy evolution.

HzRGs are known to be beacons tracing dense, proto-cluster, environments \citep[e.g.,][]{Venemans_2007}. Particularly, imaging and spectroscopic observations exploiting \textit{Spitzer}, \textit{Herschel} and the \textit{Hubble Space Telescope} (HST) have shown the robust detection of companion galaxies \citep[e.g., on cluster scales,][]{Galametz_2012,Ivison_2012,Wylezalek_2013,Wylezalek_2014,Noirot_2016,Noirot_2018}. Using MUSE/VLT, \citet{Vernet2017} and \citet{wang2023} found evidence that the $\sim100\,$kpc-scale emission halos around HzRGs are ``contaminated" by companions, which also supports this scenario. These works provided information about early clustering and AGN feedback on the scale of circumgalactic medium (CGM) and beyond. Due to the technical constraints, studies of the environments around HzRGs so far are limited to the CGM. Several works have looked into well-known HzRGs with starbursts using primarily observations from ALMA. The Dragonfly Galaxy (\object{MRC 0152-209} at $z=1.92$) is found to be going through a major merger process with a gas rich companion \citep{Emonts_2015, ZhongYuxing_2024, Lebowitz_2023}. \object{4C+41.17}, at $z=3.8$, also shows evidence of a merger with a molecular stream feeding star formation \citep{Rocca-Volmerange_2013, Emonts_2023}.  \citet{Perez-Martinez_2024} and \citet{Shimakawa_2024} studied the famous HzRG proto-cluster, the Spiderweb at $z=2.16$, with NIRCam. So far, there are only two HzRGs with published \textit{JWST} IFU observations. On $\sim20\,$kpc scales, \citet{Wang_2024} reported a tentative detection of a companion to \object{4C+19.71} at $z=3.59$. There is also evidence of a companion to \object{TN J1338-1942} at $z=4.1$ which is known to reside in a dense environment \citep[][]{Duncan_2023,Roy_2024,Saxena_2024}. It is of great interest to inspect the new \textit{JWST} data together with resolution-matched ALMA data and study nearby companions on galactic scales for a sample of HzRGs. This is crucial to understand triggering of the most powerful cosmic jets and the evolution of their massive host galaxies \citep[$M_{\star}\sim10^{11}\,M_{\odot}$,][]{Seymour_2007,DeBreuck_2010}.

In this paper, we present the first joint analysis of \textit{JWST}/NIRSpec IFU and ALMA Band 8 observations of a sample of four HzRGs at $z\approx3.5$ and focus on the detection of nearby companions. Throughout this paper, we use the term ``companion" for an emission-line component that has either a spatial or velocity offset from the central source. In Sect. \ref{sec:data_process}, we summarize the sample properties and data processing. We present the detection methods and discuss individual companions in Sects. \ref{sec:analysis} and \ref{sec:results}, respectively. Finally, we provide an estimate of companion masses and discuss our results in Sect. \ref{sec:discussion}. Throughout this paper, we assume a flat $\Lambda$ cosmology with $H_{0} = 70\, \rm{km\,s^{-1}\,Mpc^{-1}}$ and $\Omega_{m}=0.3$. Following this cosmology, $\rm{1\,arcsec=7.25-7.32\, kpc}$ at the redshifts in this study.


\begin{table*}
 \caption{Summary of observations and properties of the HzRGs analyzed in this paper.}\label{tab:sampleobs}
 \centering
\begin{tabular}{c c c c c c c}
\hline
\hline

 Target & RA, Dec (J2000)       & Redshift  & \textit{JWST} WCS shift & ALMA beam & Stellar mass & SFR  \\
        & (hh:mm:ss, dd:mm:ss)  & $z_{\rm sys}$  &  ($\Delta \alpha$, $\Delta \delta$) & $\theta_{\rm 400GHz}$ & $\log(M_{\star}/M_{\odot})$ & $M_{\odot}\,\mathrm{yr^{-1}}$  \\
(1)     & (2) & (3) & (4) & (5) & (5) & (6) \\
\hline
TN J0121+1320 & (01:21:42.725, +13:20:58.26) & 3.5190\tablefootmark{\rm{(\dag)}}  &  ($-0.210\arcsec$, $+0.003\arcsec$)  &  $0.19\arcsec$$\times$$0.15\arcsec$  & 11.02 & $626$    \\
TN J0205+2242 & (02:05:10.676, +22:42:50.57) & 3.5060\tablefootmark{\rm{(\dag)}}  &  ($-0.122\arcsec$, $-0.181\arcsec$) & $0.14\arcsec$$\times$$0.12\arcsec$ & 10.82 & $<84$    \\
4C+03.24      & (12:45:38.377, +03:23:21.14) & 3.5657\tablefootmark{(\rm{\ddag})} &  ($-0.161\arcsec$, $-0.131\arcsec$)  & $0.23\arcsec$$\times$$0.18\arcsec$ & $<11.27$& $142$  \\
4C+19.71      & (21:44:07.512, +19:29:14.58) & 3.5892\tablefootmark{\rm{(*)}} &  ($-0.014\arcsec$, $+0.173\arcsec$)  &  $0.23\arcsec$$\times$$0.18\arcsec$   & $<11.13$ & $84$  \\ 
 \hline
\end{tabular}
\tablefoot{
(1) Name of the HzRG. (2) Coordinate of the radio core, i.e., AGN. (3) Systemic redshifts of the HzRG based on 
\tablefoottext{\rm{\dag}}{\ion{He}{ii}$1640$} or \tablefoottext{*}{ [\ion{C}{i}] if \ion{He}{ii}$1640$ is unavailable \citep[][]{kolwa2023}.} The typical uncertainty is $\pm0.0004$.  \tablefoottext{\rm{\ddag}}{The systemic redshift of \object{4C+03.24} is adopted as the [\ion{O}{iii}]$5007$ redshift from \citet{Nesvadba_2017b} based on VLT/SINFONI data. Its $z_{\rm [\ion{C}{i}]}=3.5828$ may not trace the host galaxy of radio AGN (see Sect. \ref{subsec:comp_4c03}).} (4) WCS correction of the NIRSpec IFU data cube. (5) Size and position angle of the synthesized beam of the cleaned ALMA cube combined with two consecutive spectral windows. Robust = +0.5 Briggs weighting was applied for TN J0121+1320 and 4C+19.71 and robust = +2.0 Briggs weighting was applied for TN J0205+2242 and 4C+03.24. (6) Stellar mass, $M_{\star}$, and SFR from \citet{DeBreuck_2010} and \citet{Falkendal_2019} based on SED fitting.
}
\end{table*}

\section{Sample, observations and data processing} \label{sec:data_process}

\subsection{HzRGs sample}\label{subsec:sample}

In this work, we search for signatures of companions and/or mergers around four HzRGs at $z\approx3.5$ through joint observations by \textit{JWST}/NIRSpec and ALMA. These HzRGs come from a group of eight at $z$$\approx$3--5 which have multi-wavelength supporting datasets \citep[e.g., \textit{Spitzer}, \textit{Herschel}, and VLT/SINFONI][]{Seymour_2007,DeBreuck_2010,Nesvadba_2007a,Nesvadba_2017b}. In particular, they have  observations from MUSE and ALMA Band 3 focusing on the ionized CGM gas (in rest frame UV) and molecular gas \citep{wang2023,kolwa2023}. Our new NIRSpec IFU and new ALMA Band 8 observations cover rest frame optical and FIR in three-dimensional (3D) data cubes at matched sub-kpc resolution.

Besides the prerequisite selection for multi-wavelength coverage in three dimensions, these four HzRGs also have a wide range of host properties and radio morphologies. Their host galaxies are very massive with $M_{\star}\sim10^{11}\,M_{\odot}$. Their SFR span a wide range from upper limit of $84\,M_{\odot}\,\mathrm{yr^{-1}}$ (\object{TN J0205+2242}) to  $\sim 626 \,M_{\odot}\,\mathrm{yr^{-1}}$ (\object{TN J0121+1320}, Table \ref{tab:sampleobs}). \object{TN J0121+1320} is found to be on the $M_{\star}-$SFR main sequence, while the others show signatures of quenching \citep[i.e., below star-forming main sequence and having low gas content,][]{Falkendal_2019,kolwa2023}. \object{TN J0121+1320} is also the only source with an unresolved radio morphology. The others all have double-lobed morphologies. \object{4C+19.71} has the most extended jet, with two hot spots separated by $\sim60\,$kpc \citep[e.g.,][]{Pentericci_1999}. \object{4C+03.24} shows evidence of a jet being deflected \citep[e.g.,][]{vanOjik_1996}. This variety in radio morphology implies differences in terms of jet age, range over which jet-cloud interactions can take place, and etc.

The systemic redshifts of the sample used in this work are presented in Table \ref{tab:sampleobs}. Except for \object{4C+03.24}, the systemic redshifts are derived from \ion{He}{ii}1640 or [\ion{C}{i}](1-0) ([\ion{C}{i}] hereafter) emission at the radio core \citep[][]{kolwa2023}. The [\ion{C}{i}] emission of \object{4C+03.24} was known to have $ \Delta v_{[\ion{C}{i}]}=1123\pm282\,\rm km\,s^{-1}$ redshifted from the warm ionized gas prior to \textit{JWST} observations \citep[e.g.,][]{wang2023}. This is confirmed by our new NIRSpec IFU data. Throughout this work, we adopt the redshift determined from [\ion{O}{iii}]$5007$ based on VLT/SINFONI data \citep{Nesvadba_2017b}. These redshifts are from observations with lower spatial resolution and sensitivity than the data presented here. They are assumed to trace diffused gas ionized by the AGN and are found typically within $\sim300\,\rm km\,s^{-1}$ of the core [\ion{O}{iii}]$5007$ velocity measured by NIRSpec IFU.

We note that there was no prerequisite for any of the HzRGs to be in cluster environments on large scales. The only common features of these four are their similar redshifts, multi-wavelength supporting data, and bright Ly$\alpha$ emission nebulae. 

\begin{figure*}
    \centering
    \includegraphics[width=\textwidth,clip]{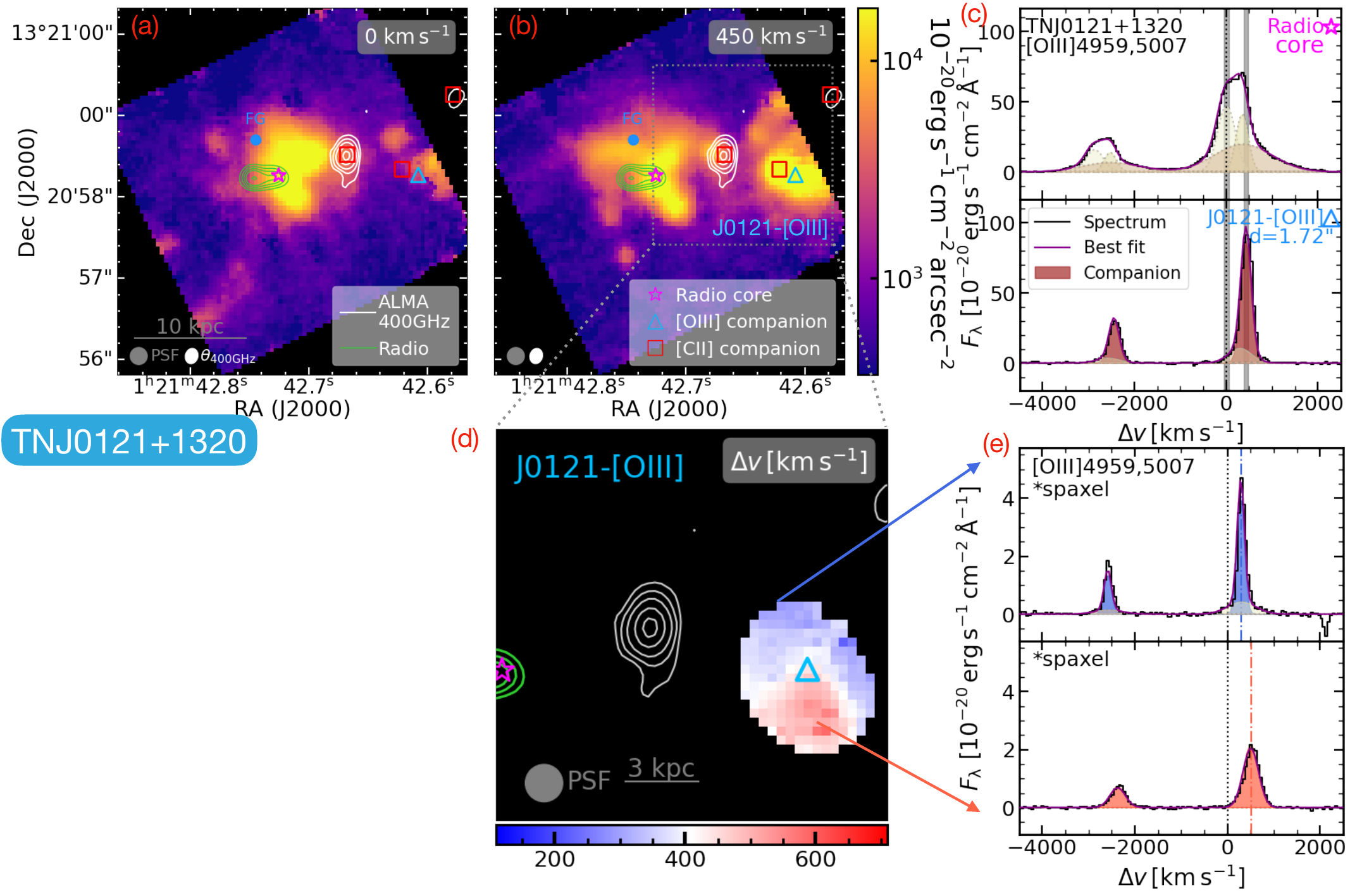}
    \caption{Emission line maps and spectra of \object{TN J0121+1320} showing the [\ion{O}{iii}] companion search. \textit{(a, b)}: [\ion{O}{iii}]$5007$ channel maps. Each map is integrated over three consecutive wavelength channels (i.e., $\sim150\,\rm km\,s^{-1}$). The central velocity is marked at the top right corner. We mark the position of the radio AGN with a magenta star. The position of [\ion{O}{iii}] companion and the [\ion{C}{ii}] companions are marked by a blue triangle and red square, respectively. The position of the foreground galaxy is marked with a blue dot. The ALMA 400$\,$GHz continuum is shown as white contours (levels of [0.35, 0.63, 1.1, 2.0, 3.0]$\,\rm mJy\,beam^{-1}$). The $8\,$GHz radio emission is shown as green contours. We illustrate the sizes of the NIRSpec IFU PSF (gray) and the synthesized beam of ALMA 400$\,$GHz ($\theta_{\rm 400GHz}$ white) at the bottom left of each panel. \textit{(c)}: [\ion{O}{iii}]$4959,5007$ spectra at the positions of the radio AGN (upper) and J0121-[\ion{O}{iii}] (lower). These continuum-subtracted spectra were extracted using $r=0.1\arcsec$ apertures. The best fits from \texttt{q3dfit} are in dark magenta. Different Gaussian components are shown in shaded colors. The fitted Gaussian component of J0121-[\ion{O}{iii}] is highlighted in dark red. The vertical shaded regions mark the velocity ranges where the channel maps in panels (a, b) are integrated. The projected distance between the [\ion{O}{iii}] companion and radio core is marked at the top right corner of the lower panel.  \textit{(d)}: Velocity map of J0121-[\ion{O}{iii}] constructed from the \texttt{q3dfit} fitting. The markers and contours are the same as in panels (a, b). \textit{(e)}: Example [\ion{O}{iii}]$4959, 5007$ spectra from two spaxels showcasing the different kinematics across J0121-[\ion{O}{iii}]. The Gaussian components are highlighted in blue and red for bluer and redder $\Delta v$, respectively.
    }
    \label{fig:O3_chanmap_0121}
\end{figure*}

\begin{figure*}
    \centering
    \includegraphics[width=\textwidth,clip]{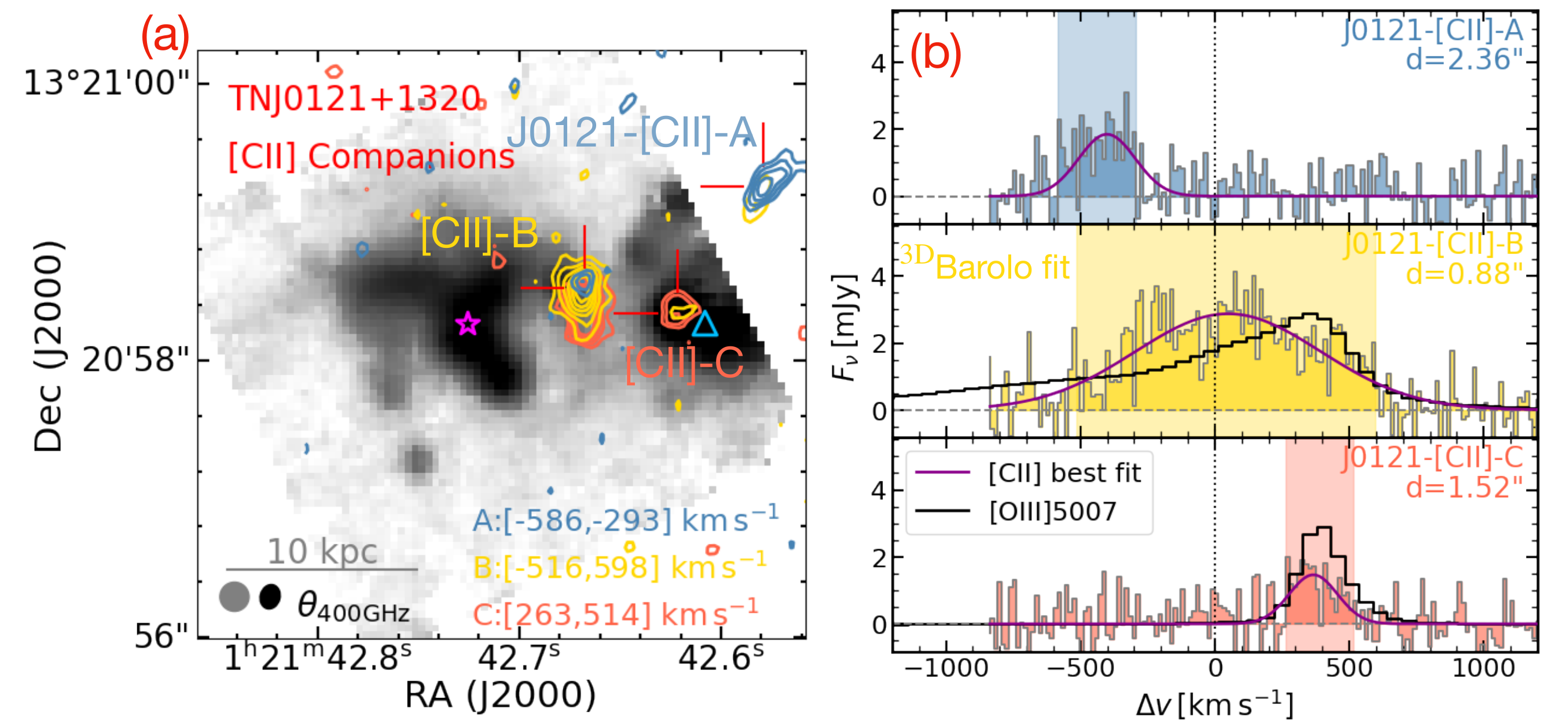}
    \caption{Emission line maps and spectra of \object{TN J0121+1320} showing evidence of [\ion{C}{ii}] companion detection.  \textit{(a)}: [\ion{C}{ii}] moment 0 maps in contours integrated at different channels as indicated in the bottom right corner. The contours are given in [3, 4, 5, ...]$\times\sigma$ levels, where $\sigma$ is the rms noise of the map. We note that only odd $\sigma$ levels are used for [\ion{C}{ii}]-B for better visualization. The gray scale background image is the same as Fig. \ref{fig:O3_chanmap_0121}b. The red open crosses mark the positions of the [\ion{C}{ii}] companions, i.e., the same as the red boxes in Fig. \ref{fig:O3_chanmap_0121}. \textit{(b)}: [\ion{C}{ii}] spectra extracted at the [\ion{C}{ii}] companion positions. The vertical shaded regions mark the velocity ranges where the moment 0 maps in panel (a) are integrated. The single Gaussian fit of [\ion{C}{ii}] spectra are shown as dark magenta curves. Where available, we show the [\ion{O}{iii}]5007 spectra in black in arbitrary units extracted at the same position with the same aperture as the [\ion{C}{ii}] companion. We mark the projected distance between the [\ion{C}{ii}] and the radio core at the top right corners.}
    \label{fig:j0121_c2_map_sp}
\end{figure*}

\begin{figure*}
    \centering
    \includegraphics[width=\textwidth,clip]{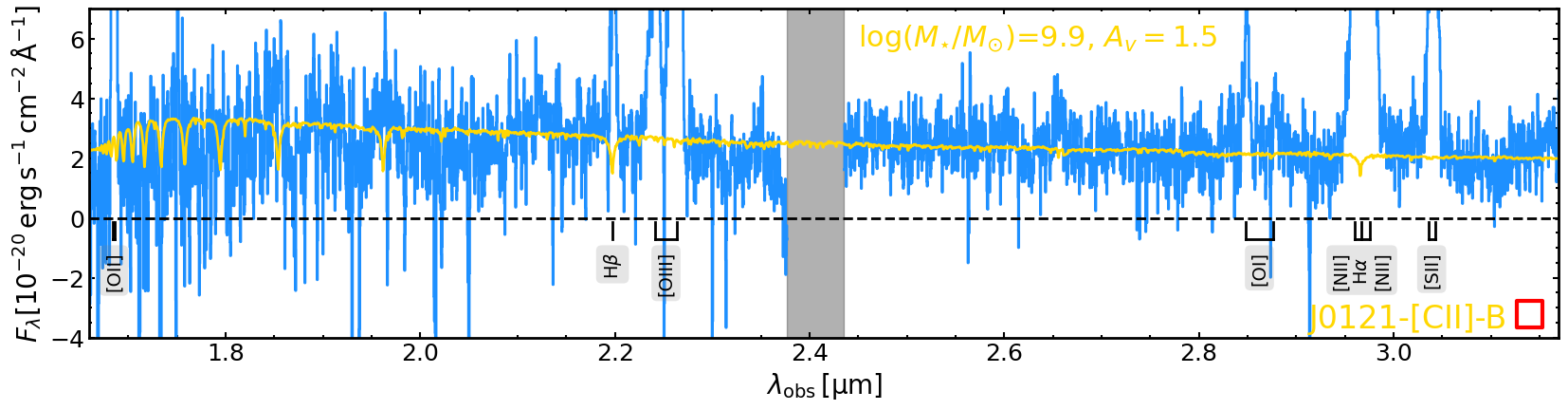}
    \caption{NIRSpec IFU spectrum of J0121-[\ion{C}{ii}]-B extracted from $r=0.2\arcsec$ aperture to maximize the continuum. The zoom-in focuses on the continuum shape. We mark the positions of bright emission lines. The yellow spectrum is the Bagpipe model with $M_{\star}=10^{9.9}\,M_{\odot}$ and $A_{V}=1.5$ \citep{Carnall_2018_Bagpipe}. The vertical gray shaded region indicates the detector gap.}\label{fig:sp_0121_continuum_opt}
\end{figure*}

\begin{figure*}
    \centering
    \includegraphics[width=\textwidth,clip]{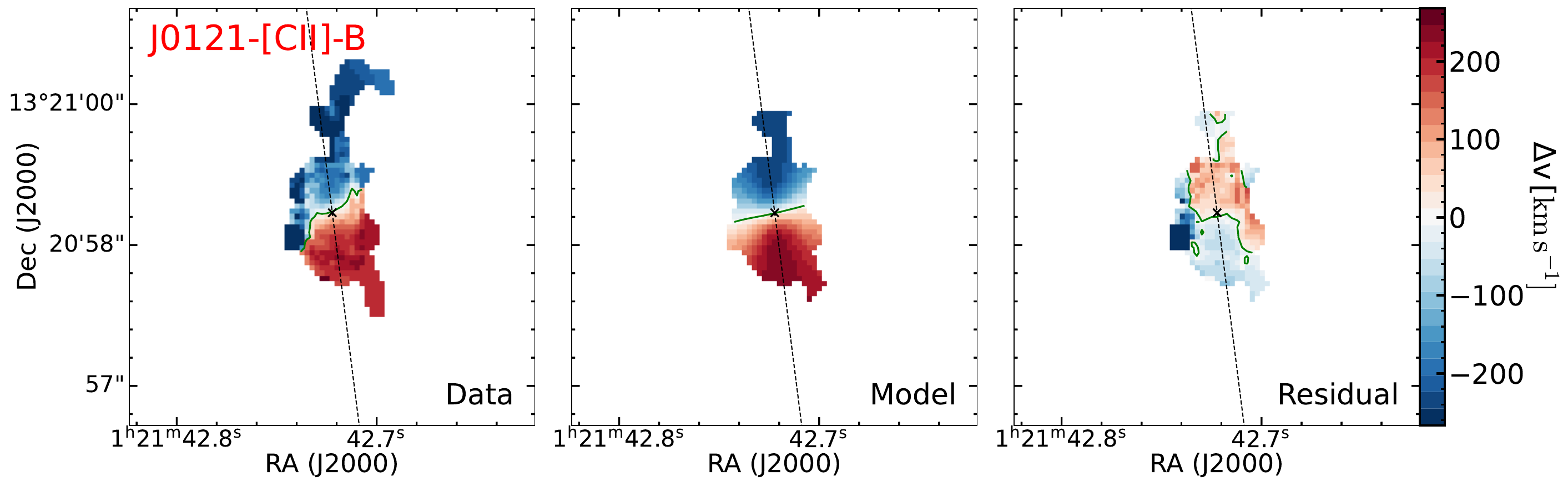}
    \caption{Moment 1 (line-of-sight velocity) map of $^{\rm 3D}$\textsc{Barolo} [\ion{C}{ii}] fit to J0121-[\ion{C}{ii}]-B. The data, model, and residual are shown in the left, middle, and right panels, respectively. The black cross marks the kinematic center. The black dashed line indicates the kinematic major axis. The green contours mark zero velocity, which has a $\Delta v=114\,\rm km\,s^{-1}$ offset between the systemic redshift (Table \ref{tab:sampleobs}).}
    \label{fig:barolo_m1_0121}
\end{figure*}

\begin{figure*}
    \centering
    \includegraphics[width=\textwidth,clip]{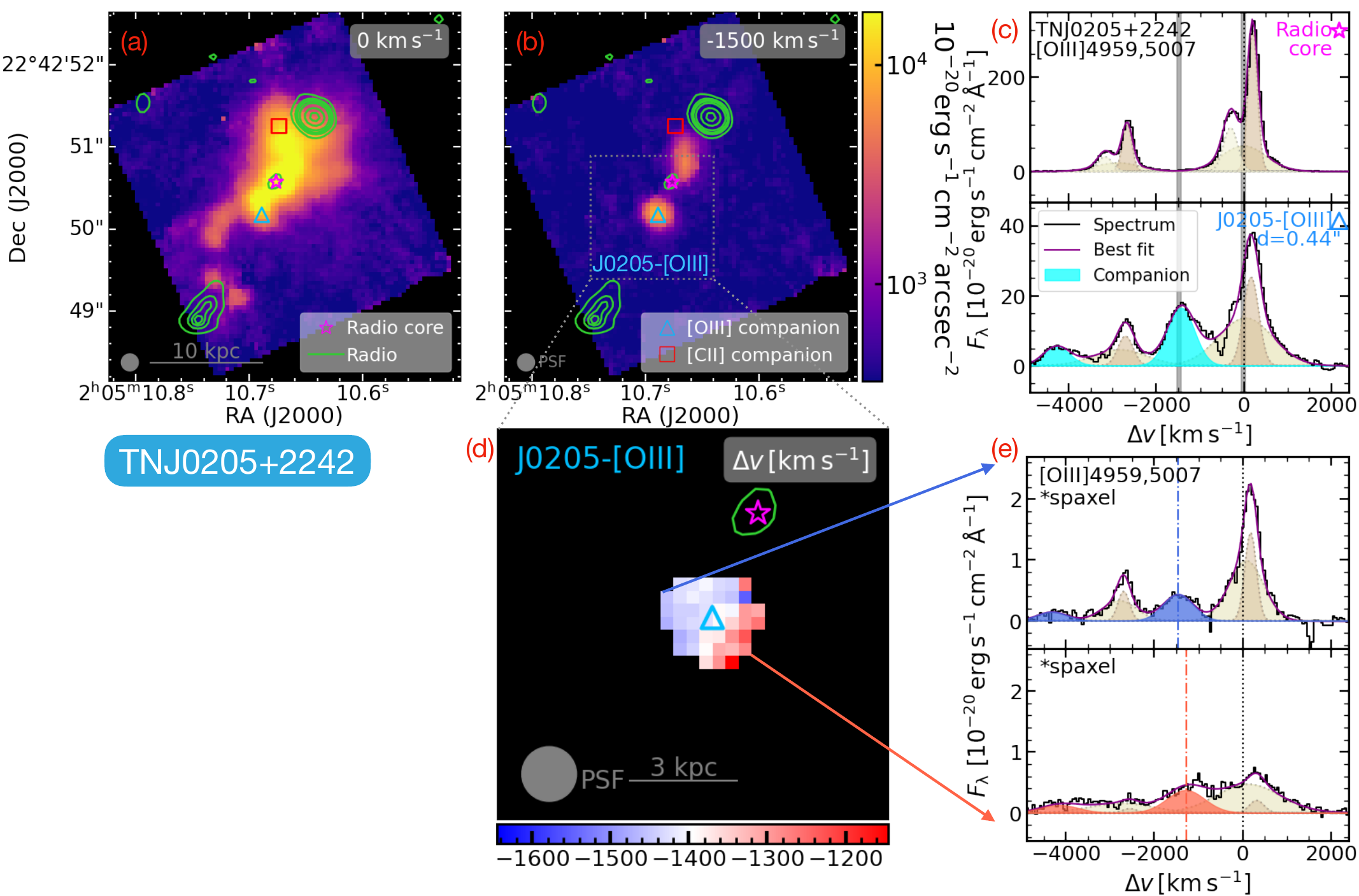}
    \caption{Similar as Fig. \ref{fig:O3_chanmap_0121}, emission line maps and spectra of \object{TN J0205+2242} showing the [\ion{O}{iii}] companion search. The northern [\ion{O}{iii}]5007 emission blob in panel (b) could be an outflowing component spectrally related to radio core. The Gaussian fit to [\ion{O}{iii}] for J0205-[\ion{O}{iii}] is highlighted in cyan in panel (c). We note that spectra in panel (c) have different y-axis limits.
    }

    \label{fig:O3_chanmap_0205}
\end{figure*}

\begin{figure*}
    \centering
    \includegraphics[width=\textwidth,clip]{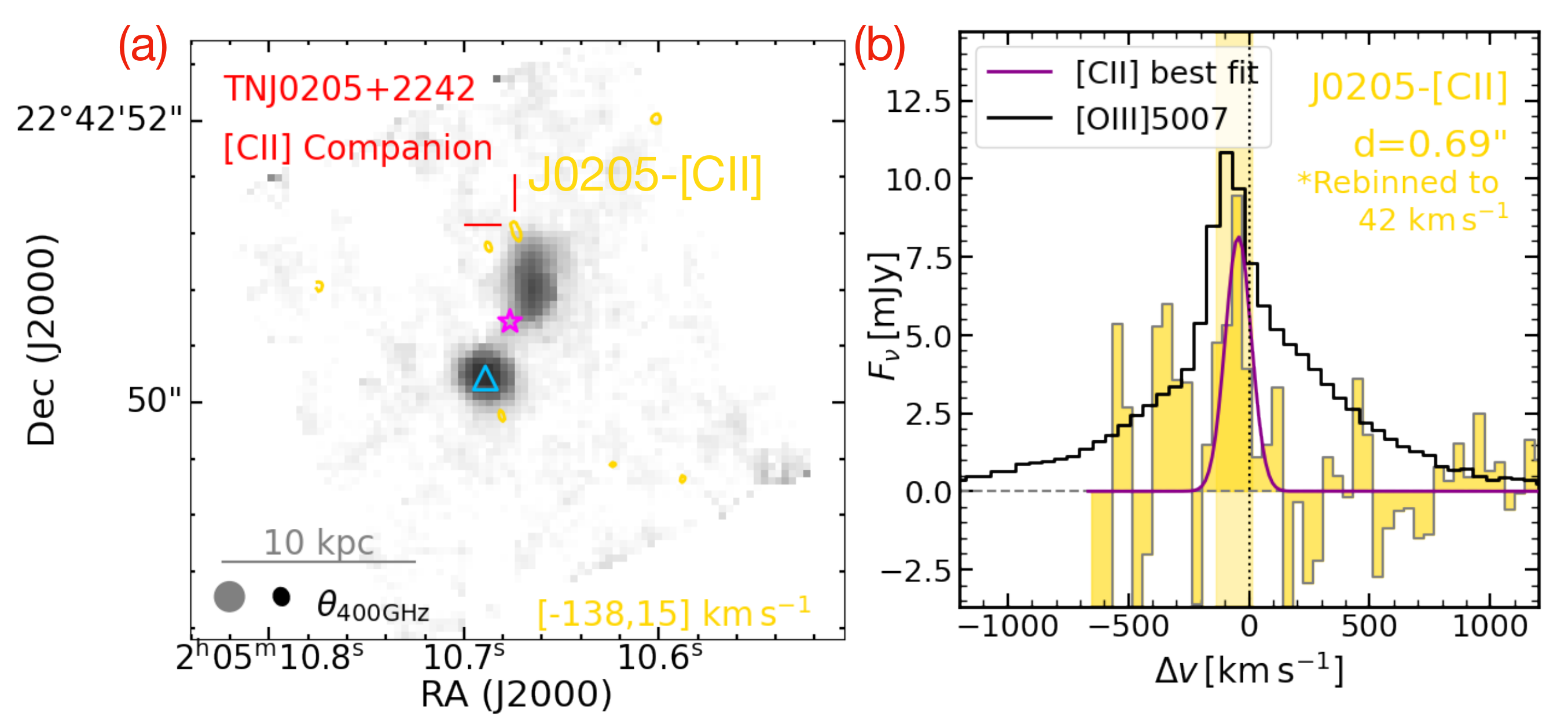}
    \caption{Similar as Fig. \ref{fig:j0121_c2_map_sp}, [\ion{C}{ii}] moment 0 map and spectrum of J0205-[\ion{C}{ii}]. Given the low S/N, the [\ion{C}{ii}] spectrum has been rebinned by a factor of three to $42\,\rm km\, s^{-1}$. }
    \label{fig:j0205_c2_map_sp}
\end{figure*}

\begin{figure*}
    \centering
    \includegraphics[width=\textwidth,clip]{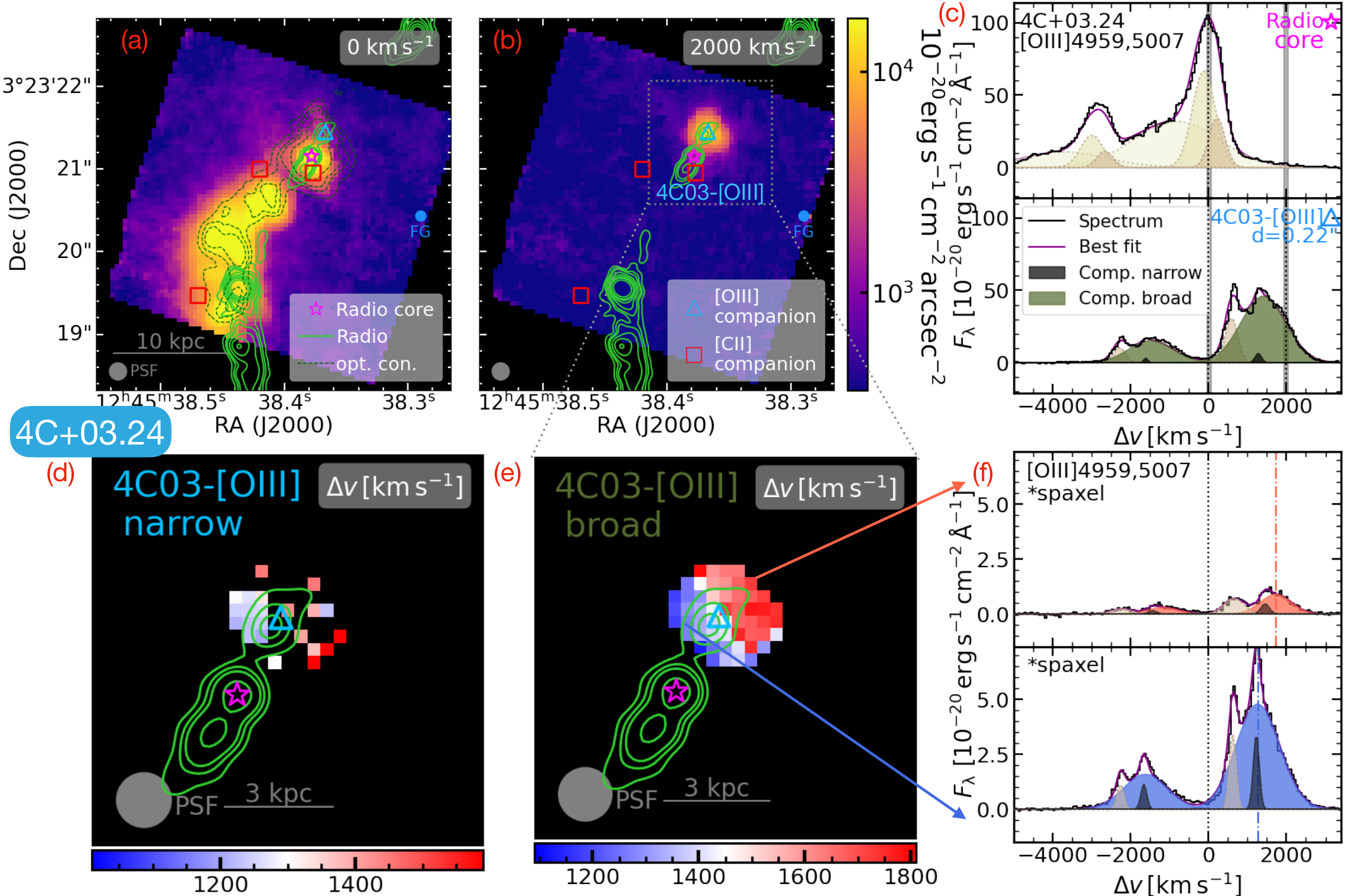}
    \caption{
    Similar as Fig. \ref{fig:O3_chanmap_0121}, emission line maps and spectra of \object{4C+03.24} showing the [\ion{O}{iii}] companion search. The dashed contours in panel (a) show the optical continuum from NIRSpec IFU. We note that we mask the continuum from the foreground galaxy (blue dot, Appendix \ref{app:4C03_fore}). There are two [\ion{O}{iii}] Gaussian components which may be [\ion{O}{iii}] companions (panel c lower). We highlight them in dark green and black for the broad and narrow components, respectively. The fitted velocity maps of these two components of 4C03-[\ion{O}{iii}] are shown in panel (d) and panel (e) for the narrow and broad components, respectively. Example spectra at two spaxels are shown in panel (f) with the broad component highlighted in red and blue.
    }

    \label{fig:O3_chanmap_4c03}

\end{figure*}

\begin{figure*}
    \centering
    \includegraphics[width=\textwidth,clip]{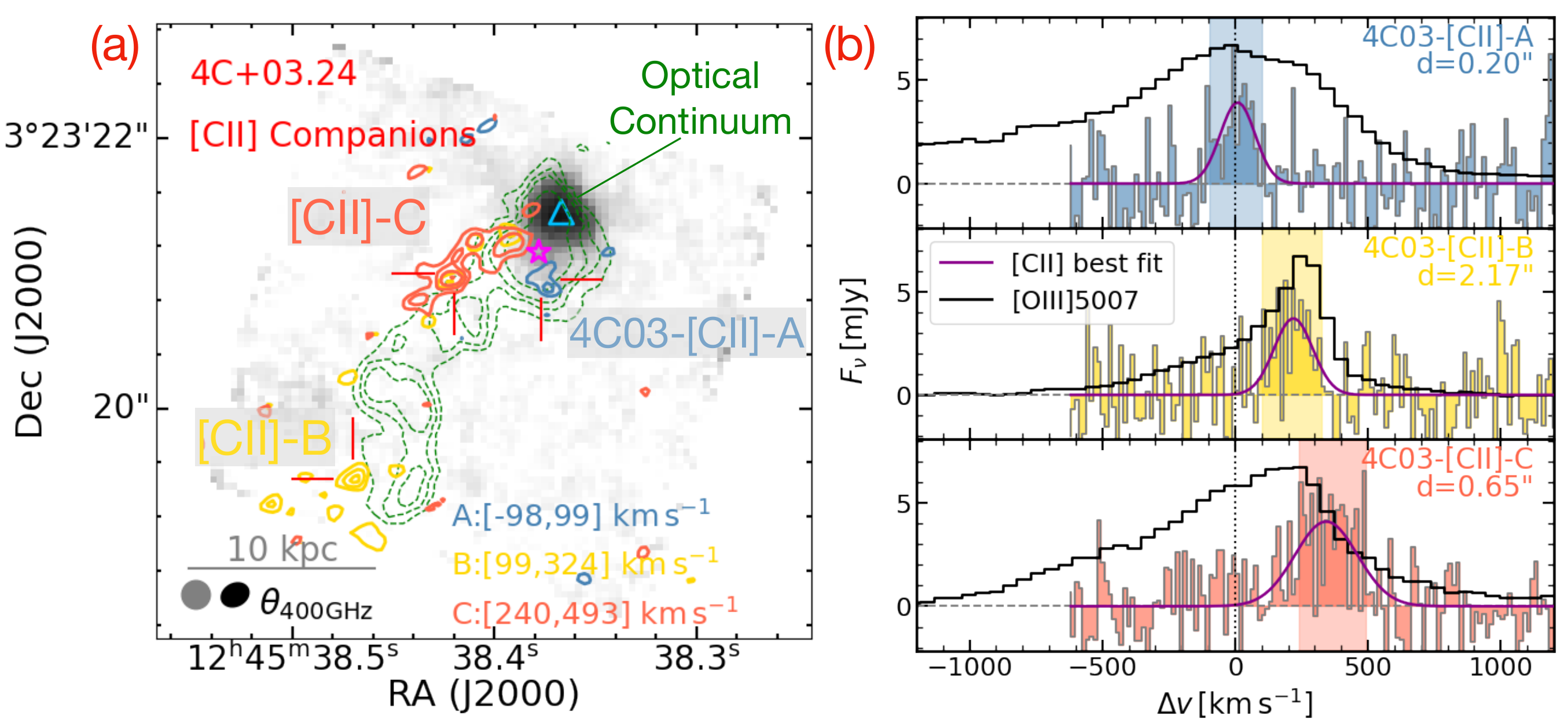}
    \caption{Similar as Fig. \ref{fig:j0121_c2_map_sp}, [\ion{C}{ii}] moment 0 maps and spectra of 4C03-[\ion{C}{ii}]. 4C03-[\ion{C}{ii}]-A is at the radio core. 4C03-[\ion{C}{ii}]-C is elongated and its position, marked with a red open cross, is marked at its $5\sigma$ peak. We overlay the NIRSpec IFU continuum with dashed green contours (the same as in Fig. \ref{fig:O3_chanmap_4c03}a) which may be the stellar tidal tail of a stripped companion galaxy. }
    \label{fig:4c03_c2_map_sp}
\end{figure*}

\begin{figure*}
    \centering
    \includegraphics[width=\textwidth,clip]{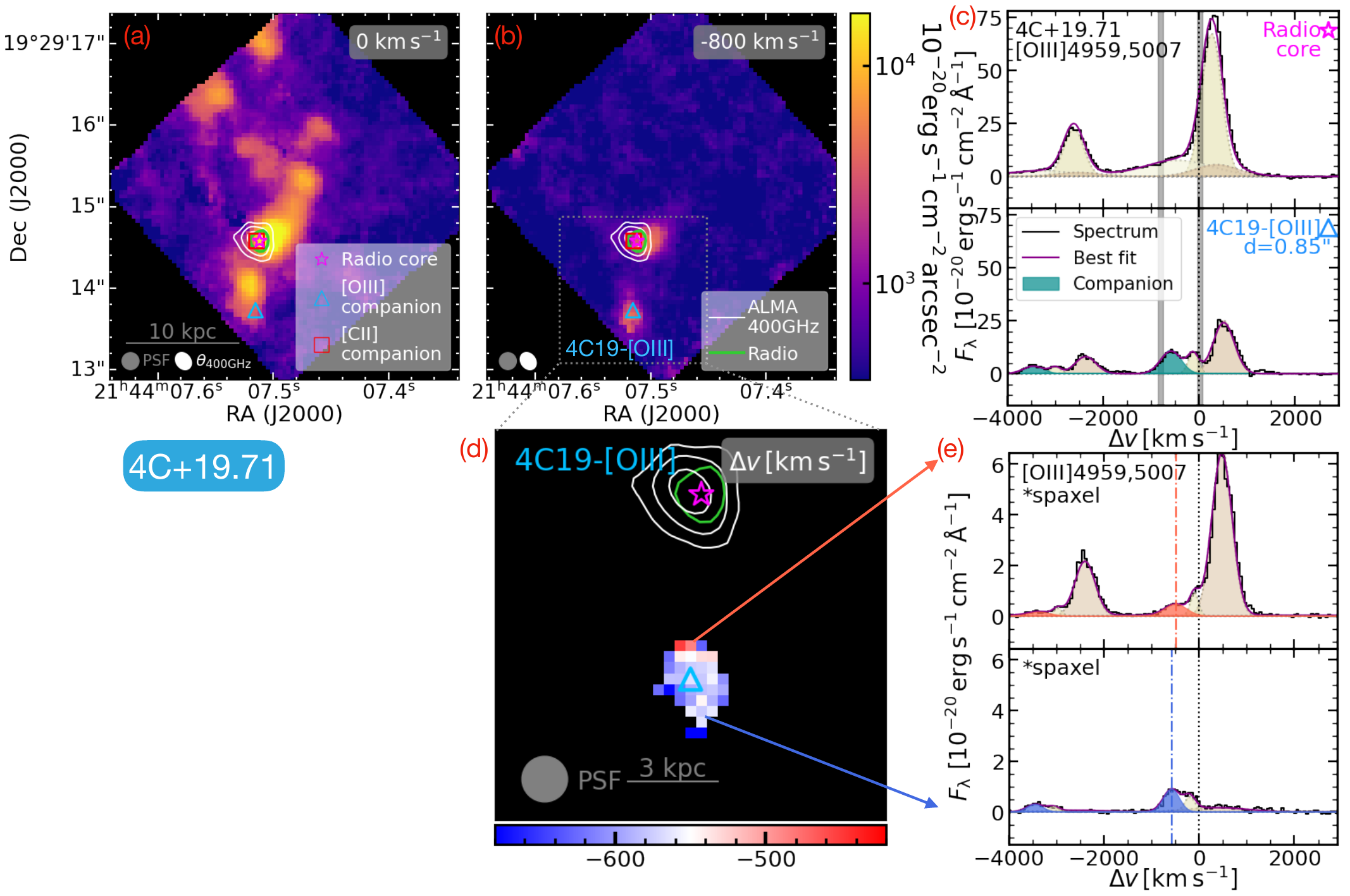}
    \caption{
    Similar as Fig. \ref{fig:O3_chanmap_0121}, emission line maps and spectra of \object{ 4C+19.71} showing [\ion{O}{iii}] companion search. The fitted [\ion{O}{iii}] Gaussian component of 4C19-[\ion{O}{iii}] is highlighted in teal in panel (c).
    }

    \label{fig:O3_chanmap_4c19}

\end{figure*}

\begin{figure*}
    \centering
    \includegraphics[width=\textwidth,clip]{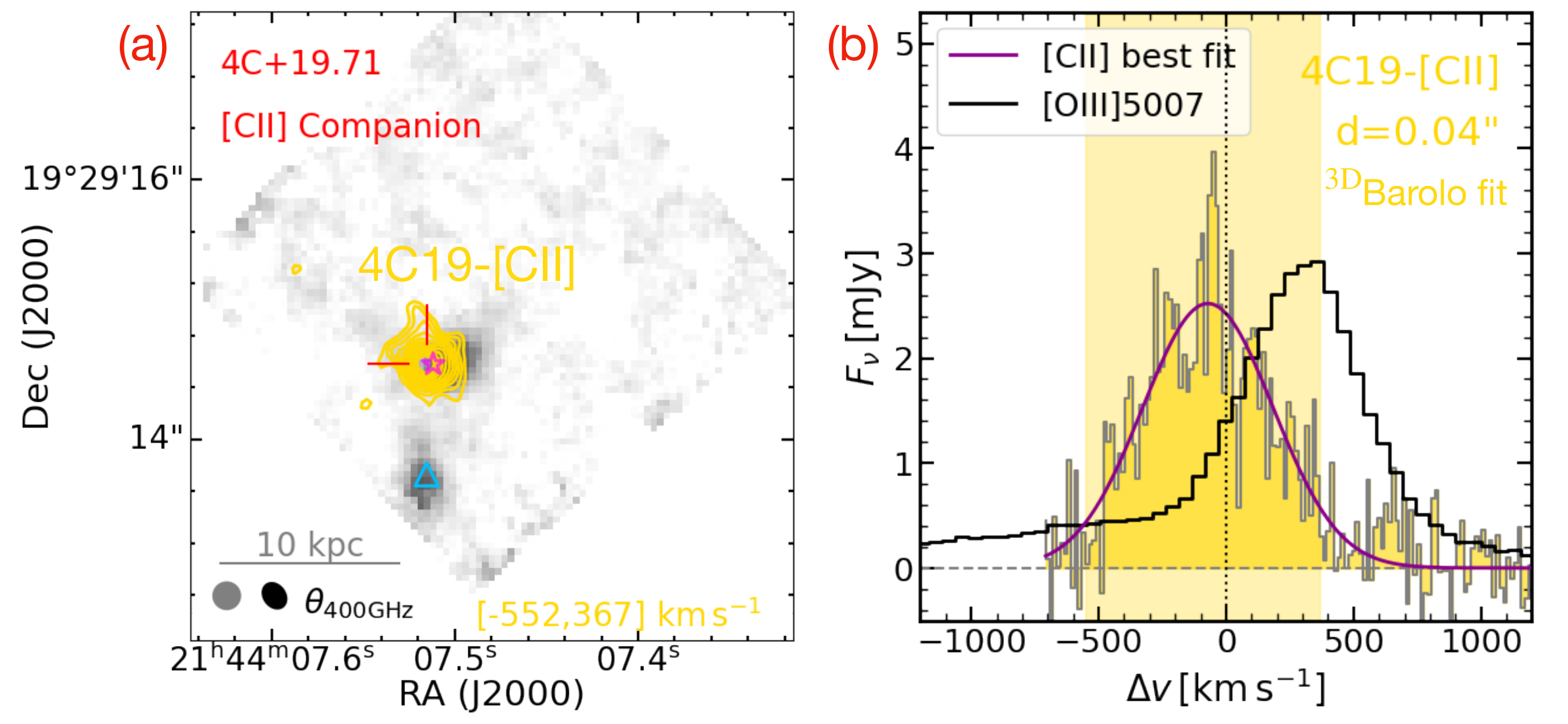}
    \caption{
    Similar as Fig. \ref{fig:j0121_c2_map_sp}, [\ion{C}{ii}] moment 0 map and spectrum of 4C19-[\ion{C}{ii}]. The right panel highlights the velocity difference between [\ion{C}{ii}] and [\ion{O}{iii}].
    }

    \label{fig:4c19_c2_map_sp}
\end{figure*}

\begin{figure*}
    \centering
    \includegraphics[width=\textwidth,clip]{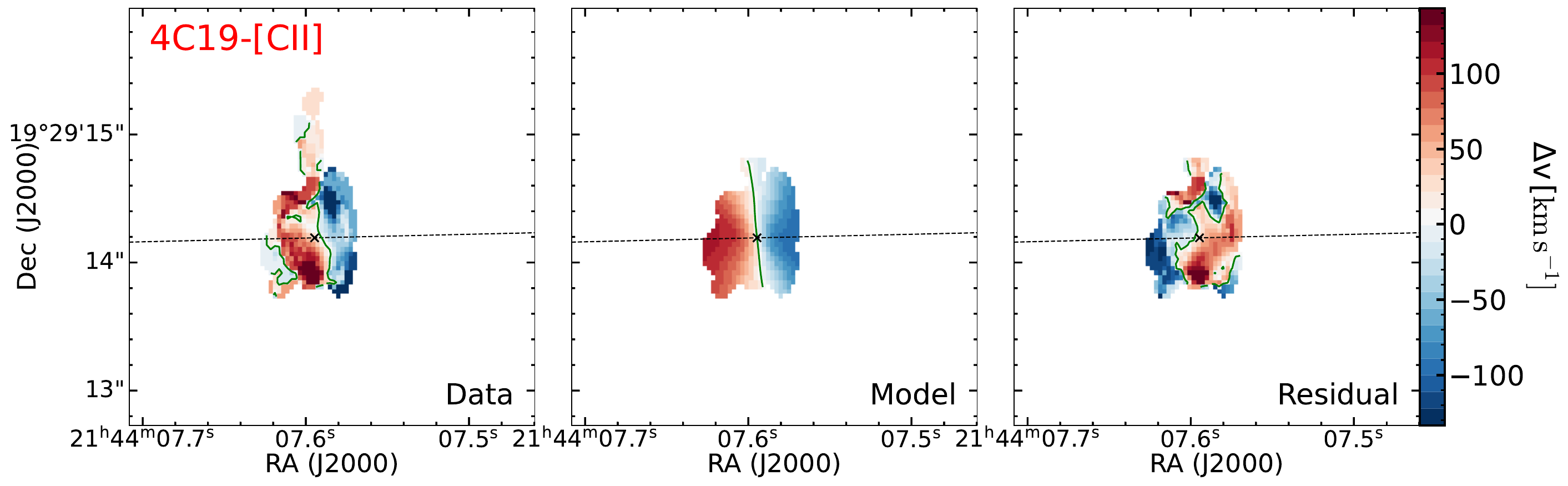}
    \caption{Similar as Fig. \ref{fig:barolo_m1_0121}, $^{\rm 3D}$\textsc{Barolo} fit moment 1 maps of 4C19-[\ion{C}{ii}]. There is a $\Delta v=-70\,\rm km\,s^{-1}$ offset between the zero velocity fitted by $^{\rm 3D}$\textsc{Barolo} and $z_{\rm sys}$ (Table \ref{tab:sampleobs}). This is consistent with the single Gaussian fit (Table \ref{tab:1d_c2_spec_fit}).}
    \label{fig:barolo_m1_4c19}
\end{figure*}

\subsection{JWST/NIRSpec IFU}\label{subsec:jwst_obs}

The NIRSpec IFU data presented here were taken from the \textit{JWST} Cycle 1 General Observer (GO) program JWST-GO-1970 (PI: Wuji Wang) between UT 2022 October 30 and 2023 August 29. Our observations targeted the four HzRGs (Sect. \ref{subsec:sample}) at $z\approx3.5$ using G235H/F170LP. This offers the wavelength coverage of $1.70-3.15\,\mathrm{\mu m}$ in the observed frame or $\sim3700-7620\,\mathrm{\AA}$ in the rest frame with a spectral resolution of $85-150\,\mathrm{km\,s^{-1}}$. In this setup, we capture all major optical emission lines seen in type-2 AGN (i.e., [\ion{O}{ii}]$3726,\,3729$ to H$\alpha$, including [\ion{S}{ii}]$6716,6731$) at once \citep[e.g.,][]{Zakamska_2003}. We adopted a 9-point dither pattern and the improved reference sampling and subtraction (IRS$^2$) read-out pattern. The total on-source exposure time ranged from  3.7 to 4.0~h per target. For \object{4C+03.24} and \object{4C+19.71}, we included the leakage exposure, $\sim0.4$~h each, to identify contamination light due to failed open shutters on the microshutter array (MSA) at the first dither position. For \object{TN J0205+2242} and \object{TN J0121+1320}, we designed pointing verification imaging exposures, $\sim233$~s each, after the science observation with MSA shutters all-open to improve the astrometry \citep[e.g.,][]{Wang_2024}.

We downloaded all raw data from the Mikulski Archive for Space Telescopes (MAST). We used v1.13.4 of the {\em JWST} Science Calibration pipeline\footnote{\url{https://github.com/spacetelescope/jwst}} with the Calibration Reference Data System (CRDS) context file jwst\_1202.pmap. Our procedure is similar to \citet{Vayner_2023Q3D} and \citet{Wang_2024}. Specifically, we executed the standard first and second stages of the pipeline. We subtracted the $1/f$ noise from the rate files resulted from stage 1. The stage 3 of the standard pipeline does not perform well the outlier rejection. Therefore, we used sigma clipping to combine the dithered exposures into the final data cube. The primary source of outliers for our observations is likely cosmic rays due to the adopted readout mode. We did not include flux calibration step as in \citet{Wang_2024} since the standard pipeline already has a good performance (i.e., relative flux difference $<5\%$ for \object{4C+19.71}). The flux density unit of the final cube was converted into $\mathrm{erg\,s^{-1}\,cm^{-2}\,\AA^{-1}}$. 

Through out the field of view (FoV), there is an uniform background continuum emission with a flux density of several $\mathrm{10^{-21}\,erg\,s^{-1}\,cm^{-2}\,\AA^{-1}}$, and it is brighter at shorter wavelength. To subtract this background, we first extracted the median background spectrum by spatially masking the continuum emission sources and continuum-like patterns (presumably leakage emission from the MSA). Then, we fitted a fifth or sixth order polynomial to the median background spectrum. Finally, we subtracted this fitted contribution from the whole data cube. The final cubes have a rms noise at 2.6$\,\rm \mu m$ of $\sim4$ to $6\times \mathrm{10^{-22}\,erg\,s^{-1}\,cm^{-2}\,\AA^{-1}}$  per wavelength channel. 

The point spread function (PSF) of the NIRSpec IFU depends on wavelength and spatial position. The HzRGs analyzed in the paper are all type-2 AGN, meaning the presumed bright point-like quasar is obscured. Hence, it is not necessary for us to subtract the PSF of the quasar. The full width at half maximum (FWHM) of the \textit{JWST} PSF used in this work is given as the one constructed in \citet{Vayner_2023Q3D_2}, which has a typical size of $\sim0.11\arcsec$.

\subsection{ALMA Band 8}\label{subsec:alma_obs}
Our ALMA Band 8 observations were taken under the program ID 2021.1.00576S (PI: Wuji Wang). The observations of \object{TN J0121+1320} and \object{4C+19.71} were fully carried out. For the remaining two, only half of the proposed integration times were executed. Despite the shallow exposure time, [\ion{C}{ii}] lines are detected in the field of \object{TN J0205+2242} and \object{4C+03.24}. However, the FIR continuum of these two are undetected. These observations were carried out in C43-5 configuration between UT 2022 June 17 and 2022 August 10. We set up one spectral window in Band 8 to be centered at the redshifted frequency of [\ion{C}{ii}] ($\nu_{\rm rest}=1900.537\,$GHz). Then, the other three spectral windows were used to observe dust emission. We note that the unexpected broad [\ion{C}{ii}] line width relative to the width of spectral window (1.875$\,$GHz $\sim$ 1350$\,\rm km\,s^{-1}$) and the redshift uncertainty end up of the line detection in two consecutive spectral windows, especially \object{TN J0121+1320} and \object{4C+19.71} (see processing below). The default ALMA bandpass and phase calibrations were adopted.

We processed the data using the Common Astronomy Software Applications for Radio Astronomy v6.6.0 \citep[CASA,][]{CASA_2022}. The task \texttt{tclean} was used to clean the cube down to a $2\sigma$ level. For \object{TN J0121+1320} and \object{4C+19.71}, we use the cubes with Briggs weighting of robust$=+0.5$ in the analysis. For \object{TN J0205+2242} and \object{4C+03.24}, we use the cubes with Briggs weighting of robust$=+2.0$. This provides higher signal to noise ratio (S/N) with a trade off in spatial resolution. We cleaned the two consecutive spectral windows with [\ion{C}{ii}] together without continuum subtraction in the $uv-$plane. The shapes of the spectra at the transition frequencies of the two windows were compared to the cubes cleaned individually. We did not find any jump in flux. Throughout the paper, we used the cubes and continuum maps (for the two with detections) after the primary beam correction. The final cubes were smoothed to $14,\rm km,s^{-1}$ per channel. The choice of resolution element is a balance among S/N, continuum coverage and a sufficient number of channels for dynamical study (especially for TN J0121+1320 and 4C+19.71). The rms noise of the [\ion{C}{ii}] cubes at the systemic redshift are $3.0\,\rm mJy\,beam^{-1}$, $1.3\,\rm mJy\,beam^{-1}$, $1.1\,\rm mJy\,beam^{-1}$, and $1.0\,\rm mJy\,beam^{-1}$ per channel for \object{TN J0205+2242}, \object{TN J0121+1320}, \object{4C+03.24}, and \object{4C+19.71}, respectively. The rms noise for the continuum maps constructed from the  individual spectral windows are $76\,\rm \mu Jy\,beam^{-1}$ and $92\,\rm \mu Jy\,beam^{-1}$ for \object{TN J0121+1320} and \object{4C+19.71}, respectively. The continuum images for these two targets from the spectra windows centered at the lowest frequency, i.e., separated the most from the windows with line emission. For the remaining two without FIR continuum detections, we report $3\sigma$ upper limits of $0.28\,\rm mJy\,beam^{-1}$ and $0.25\,\rm mJy\,beam^{-1}$ for \object{TN J0205+2242} and \object{4C+03.24}, respectively. These upper limits were calculated from the continuum images constructed using all four spectral windows.

\subsection{Astrometric alignment of NIRSpec IFU cubes}\label{subsec:WCS_align}

The absolute astrometric calibration of NIRSpec IFU observations is known to have a $\sim0.2-0.3\arcsec$ uncertainty \citep[][and private communication with NIRSpec instrument scientists at STScI]{Perna_2023,Wang_2024}. Hence, we manually align the World Coordinate System (WCS) of our targets to other datasets with known WCS. Our manual shifts are summarized in Table \ref{tab:sampleobs}. The assumption behind this alignment is that the continuum detected by the NIRSpec IFU is at the position of the radio AGN (i.e., core position of the radio jet). The NIRSpec IFU continuum may be affected by leakage light through microshutter array and other artifacts. We checked the continuum and found that the continuum centroids are often close, $<0.1\arcsec$, to the position of the brightness peak of the emission lines (e.g., [\ion{O}{iii}]5007), which supports our assumption. Specifically, to determine the alignment, we fitted a 2D Gaussian profile to the continuum of the radio galaxy at observed wavelength $2.6\,\rm \mu m$ (i.e., line-free region at rest frame $\sim5770\AA$ at $z=3.5$). Similarly, we fit a 2D Gaussian profile to the radio core observed at $8.4\,$GHz \citep[][]{vanOjik_1996,Nesvadba_2007a,Nesvadba_2017b}. We then shifted the WCS of the NIRSpec IFU to match the radio data.

The case of \object{TN J0121+1320} is complex. Its compact radio emission has an offset of $\sim1\arcsec$ from the ALMA Band 8 dust position. Assuming that the WCS of the two interferometers are accurate, this may already suggest an offset likely due the existence of companions. In addition, we find a foreground galaxy at $z=1.4577$ in the NIRSpec IFU cube as well as the MUSE cube whose continuum may contaminate \object{TN J0121+1320} \citep[see Appendix \ref{app:tnj0121_fore} and MUSE data in][]{wang2023}. It is located to the north of the radio emission with a projected offset of $\sim0.5\arcsec$. There is another continuum source detected by the NIRSpec IFU located west of the radio emission with an offset of $\sim0.7 \arcsec$ (discussed in Sect. \ref{subsec:comp_0121}). This continuum position is not at the same position as the brightness peak of the optical line (e.g., [\ion{O}{III}]5007). The projected separations quoted here are prior to the astrometric correction. Assuming the WCS assigned by the \textit{JWST} pipeline is accurate to $\sim0.2\arcsec$ to $0.3\arcsec$, this western continuum may not be the emission associated with the radio galaxy. Hence, our correction procedure for \object{TN J0121+1320} is based on the foreground object in the north. As its continuum is faint and presumably blended with \object{TN J0121+1320} by MUSE, we used the fitted 2D Gaussian positions of the foreground [\ion{O}{ii}]$3726,3729$ line in the MUSE data cube and its [\ion{S}{iii}]$9533$ line in NIRSpec IFU data cube to determine the WCS correction. The narrow band [\ion{O}{ii}]$3726,3729$ line emission map is spatially unresolved which leads to less uncertainty in the position determination. The resulting shift is $-0.21\arcsec$ and $0.003\arcsec$ in RA and Dec, respectively, consistent with the expected systemic uncertainty. After the correction, we found that the western optical continuum is aligned with the ALMA dust position.

In \citet{Wang_2024}, we used one of the foreground galaxies to determine the astrometric correction (G1 in their naming format and using imaging data from \textit{HST}). In this work, we updated the reduction of the $8.4\,$GHz radio observation where the previously unseen radio core is now detected at $5\sigma$ \citep[program AN129,][]{Nesvadba_2017b}. The aligned WCS of \object{4C+19.71} in this work has an offset of $\sim0.16\arcsec$ from the one in \citet{Wang_2024}. In addition, the position of the dust continuum observed by ALMA Band 8 is aligned with the radio core (Sect. \ref{subsec:alma_obs}).

\subsection{Radio maps}\label{subsec:radio_ima}
In this work, we only used the radio observations to trace the position of the jets. We adopted the VLA X-Band ($8.4\,$GHz) radio maps of \object{TN J0205+2242} published in \citet{Nesvadba_2007a}.

We used the new radio data of \object{4C+03.24} in this work \citep[see e.g.,][for previous radio data]{vanOjik_1996,wang2023}. The observation of 4C+03.24 was carried on UT 2024 December 20, UT 2025 January 02, and UT 2025 January 03 with the VLA in A configuration in the Ku-band covering a frequency range of 12 to 18 GHz (Project 24B-147, PI: Carlos De Breuck). The target was observed for 5~h (8.7~h in total) over a contiguous Local Sidereal Time (LST) range for good sampling of the $uv$-plane. The 6~GHz bandwidth was split into 48 spectral windows of 128 channels of 2~MHz width each, and a sample integration time of 2~s.  3C286 and J1224+0330 were used as the ﬂux and phase calibrators, respectively. Pointing calibration was also performed using both sources. The data were then cross-calibrated, imaged and self-calibrated in CASA. The final image was made using Briggs weighting with a \texttt{robust = 0.0}. We obtained a resolution of $0.15\arcsec\times0.13\arcsec$ (PA=56.5$^{\circ}$) and rms noise of $\sim$4.5$\rm \mu Jy\,beam^{-1}$.

We reprocessed archival VLA C-band observations of \object{4C+19.71} (\object{MG2144+1929}) using CASA and detected the previously unseen radio core \citep[][]{Nesvadba_2017b}. The observation was carried on UT 2012 December 22 with the VLA in A configuration (Project code: 12B-243). The target was observed for $\sim$15 min with an integration time of $1\,$s and 16 spectral windows of 128$\,$MHz bandwidth each, between 4.15 and 7.85$\,$GHz. For our analysis, we only used the 8 spectral windows from 6.95 to 7.85$\,$GHz. Each spectral window consisted of 32 channels of 4$\,$MHz width. 3C\,48 and J2139+1423 were used as the ﬂux and phase calibrators, respectively. The data were manually ﬂagged and cross-calibrated in CASA, set to the same ﬂux scale as above. Imaging and self-calibration were then performed using the tasks \texttt{tclean} and \texttt{gaincal} respectively. The ﬁnal image was made using Briggs weighting with robust$=+0.5$. We obtained a resolution of 0.25$\arcsec \times 0.24\arcsec$ and an rms noise of $\sim17\, \rm \mu Jy\,beam^{-1}$.

We reprocessed archival VLA X-band observations of \object{TN J0121+1320}, carried out on UT 1998 May 07 with the VLA in A configuration (Project code: AY92). The target was observed for $\sim$9$\,$min with an integration time of 10$\,$s and two spectral windows of 50$\,$MHz bandwidth centered at 8.435 and 8.485$\,$GHz each. 3C\,286 and J0121+118 were used as the ﬂux and phase calibrators, respectively. The data were manually ﬂagged and cross-calibrated using Astronomical Image Processing System \citep[AIPS,][]{Greisen_2003}. The ﬂux scale was set according to \citet{Perley_2017}. The data were then imaged using CASA. The ﬁnal image was made using task \texttt{tclean} and Briggs weighting with robust$=-0.5$. We obtained a resolution of 0.24$\arcsec \times 0.24\arcsec$ and rms noise of $\sim90\,\rm \mu Jy\,beam^{-1}$. We were not able to perform reliable self-calibration of the data due to the low S/N of the solutions. However, the source position is still reliable, which is the only information used in this analysis.


\section{Analysis} \label{sec:analysis}
In this section, we first describe general methods used to search for companions around our radio AGN. We used two independent approaches to identify companion systems: (i) peculiar kinematics of warm ionized gas observed by \textit{JWST}/NIRSpec IFU ([\ion{O}{iii}], Sect. \ref{subsec:O3_kine}) and (ii) cold gas observed by ALMA ([\ion{C}{ii}], Sect. \ref{subsec:c2_detection}). We also look for continuum emission from the companions detected by the two methods (Sect. \ref{subsec:cont_method}). We note that a systematic continuum study is beyond the scope of this work, especially optical continuum studies which would require careful treatment in data processing. Hence, in this paper, we do not leverage detection of continuum as evidence for a companion. We name the companions by their central radio AGN followed by the detection method. For example, the companion in the field of TN J0121+1320 detected through [\ion{O}{iii}] is named as J0121-[\ion{O}{iii}]. If more than one companion is detected in the same field using the same method, we name these alphabetically in the order of redshift. For example, the [\ion{C}{ii}] companion with the lowest redshift in the field of 4C+03.24 is named as 4C03-[\ion{C}{ii}]-A. There is a companion detected in both methods (i.e., J0121-[\ion{O}{iii}] and J0121-[\ion{C}{ii}]-C may be one companion, see Sect. \ref{subsec:comp_0121}). We note that our analysis may not be complete in the sense of detecting all possible companion sources in the fields of the radio galaxies, but we focus here specifically on peculiar warm ionized gas kinematics and [\ion{C}{ii}] detections. Furthermore, we deliberately refer to the detected structures as companions rather than galaxies as we do not necessarily detect stellar continuum at each location. These companions could therefore also represent isolated gas clouds, possibly related to ongoing jet activity as suggested in simulations \citep[e.g.,][]{Ondaro-Mallea_2024}.

\subsection{Warm ionized gas kinematics}\label{subsec:O3_kine}

We first search for peculiar ionized gas kinematics captured by the NIRSpec IFU as a signature of companions. [\ion{O}{iii}]$5007$ is one of the brightest optical emission lines and is a traditional kinematic tracer of warm ionized gas around AGN \citep[e.g.,][]{Harrison_2014}. Using VLT/SINFONI observations of [\ion{O}{iii}]$5007$, \citet{Nesvadba_2007a, Nesvadba_2017b} analyzed the ionized gas kinematics of a large sample HzRGs, including our targets. At a spatial resolution of $\sim0.5\arcsec$ to $0.8\arcsec$, the bulk motion of the ionized gas is found to be consistent with jet-driven outflows. \textit{JWST}/NIRSpec IFU makes it possible to examine the sub-kpc kinematics at and beyond Cosmic Noon with the same tracer \citep[e.g.,][]{Perna_2023,Vayner_2023Q3D_2,Wang_2024}. As pointed by some studies, the nature of some outflows may be confined on smaller scales, only evident at finer resolution and high sensitivity \citep[e.g.,][]{Wang_2024, Holden_2024_MUSE}. When focusing on the vicinity ($\lesssim18\,$kpc) of the brightest quasars, we often have [\ion{O}{iii}]$5007$ out to the edge of the FoV. Hence, we classify a [\ion{O}{iii}]$5007$ emitter as a companion in this study by leveraging the full 3D information of the data cube.

 Specifically, we visually inspect the [\ion{O}{iii}]$5007$ emission at different velocity channels (e.g., Fig. \ref{fig:O3_chanmap_0121}a\ref{fig:O3_chanmap_0121}b and Appendix \ref{app:o3_channel}). For each target, we detect line-emitting gas blobs of $\sim3-6\,$kpc extent which are kinematically distinct from the tens of kiloparsecs scale velocity fields seen in the SINFONI data. That is to say, these blobs are different from either large-scale outflows or gaseous nebula at the systemic redshift. To further separate the companion emission, we decompose these line-emitting components by fitting $r=0.1\arcsec$ aperture-extracted spectra using \texttt{q3dfit} \citep{Wylezalek_2022q3d,Rupke_2023q3dfit}. In \texttt{q3dfit}, the wavelength dependent spectral resolution is taken into account. The fitting is done simultaneously for all bright emission lines covered, namely [\ion{O}{ii}]$3726,3729$, H$\beta$, [\ion{O}{iii}]$4959,5007$, H$\alpha$, [\ion{N}{ii}]$6548,6583$, and [\ion{S}{ii}]$6716,6731$, while keeping the Gaussian components for different lines kinematically tied \citep[i.e., the redshift and line widths are fixed for corresponding components; see also][]{Vayner_2023Q3D,Wang_2024}. The flux ratios of doublets are set according to atomic physics: $F_{\rm [\ion{O}{iii}]5007}$/$F_{\rm [\ion{O}{iii}]4959}=3$ and $F_{\rm [\ion{N}{ii}]6583}$/$F_{\rm [\ion{N}{ii}]6548}=3$. We limit $F_{\rm [\ion{S}{ii}]6716}$/$F_{\rm [\ion{S}{ii}]6731}$ to the range [0.4375, 1.4484] and fixed the ratio of spectrally unresolved $[\ion{O}{ii}]3726,3729$ to $F_{\rm [\ion{O}{ii}]3726}$/$F_{\rm [\ion{O}{ii}]3729}=1$ \citep[e.g.,][]{Sanders_2016}. The optimized number of components are determined by \texttt{q3dfit}, with a maximum allowed number as input. For comparison, we also fit the spectra at central AGN positions through the same procedure (e.g., Fig. \ref{fig:O3_chanmap_0121}c). The fitting results are presented in Table \ref{tab:1d_o3_spec_fit}.

We further map the companion emissions spatially to demonstrate their offset from the bulk emission on larger scales. Using \texttt{q3dfit}, we fit spaxel-based spectra around the companion positions. Due to the limitation of S/N and the focus on kinematics, we limit this spatial fit only to [\ion{O}{iii}]$4959,5007$. We detect velocity gradients for three out of the four [\ion{O}{iii}] emission blobs. The gradient detection of 4C19-[\ion{O}{iii}] is tentative. We present the velocity shift maps and exemplar spectra at both ends of the gradient to demonstrate the gas kinematics (e.g., Fig. \ref{fig:O3_chanmap_0121}d\ref{fig:O3_chanmap_0121}e). We refer to these line-emitting gaseous blobs with peculiar kinematics as ``[\ion{O}{iii}] companions" and examine them individually below.

\subsection{[\ion{C}{ii}] emitters}\label{subsec:c2_detection}
The brightest FIR emission line, [\ion{C}{ii}] is a frequently used gas tracer at high-redshift \citep[e.g.,][]{Lagache_2018,Dessauges-Zavadsky_2020}. Directly associated with photodissociation regions, the detection of [\ion{C}{ii}] may indicate the presence of cold gas systems. Moreover, it has been used as a kinematic tracer of regular motion (e.g., rotation) and merger activity \citep[][]{Lelli_2021,Romano_2021,Herrera-Camus_2022}.

We examine the ALMA Band 8 data cubes and detect at least one [\ion{C}{ii}] emitters at $\gtrsim3\sigma$ in the field around each the four radio AGN. We present the [\ion{C}{ii}] moment 0 maps constructed from velocity channels where emission is detected (e.g., contours in Fig. \ref{fig:j0121_c2_map_sp}a). We then extract 1D spectra at the position of [\ion{C}{ii}] emitters (e.g., Fig. \ref{fig:j0121_c2_map_sp}b). For the field of TN J0121+1320 and 4C+19.71, we use apertures with the size of the major synthesized beam. For TN J0205+2242 and 4C+03.24, we increase the aperture to twice the size of major synthesized beam to increase the S/N.  We perform line fitting to the 1D spectra with one Gaussian component (Table \ref{tab:1d_c2_spec_fit}). For the brighter [\ion{C}{ii}] companions (S/N$>10\sigma$; i.e., J0121-[\ion{C}{ii}]-B and 4C19-[\ion{C}{ii}], see Sect. \ref{subsec:comp_0121} and \ref{subsec:comp_4c19}), we also conduct spatial fitting using $^{\rm 3D}$\textsc{Barolo} to study the dynamics of the cold ionized gas \citep[][and Appendix \ref{app:barolo}]{DiTeodoro_2015}.

We refer to these [\ion{C}{ii}] emitting systems as ``[\ion{C}{ii}] companion" and examine them individually below. To compare the warm and cold ionized gas at the position of [\ion{C}{ii}] companions, we overlay the [\ion{O}{iii}] surface brightness map and 1D spectra with the [\ion{C}{ii}] moment 0 map and spectra (e.g., Fig. \ref{fig:j0121_c2_map_sp}). Without further information, we cannot rule out the possibility that the ``[\ion{C}{ii}] emissions" without corresponding [\ion{O}{iii}] emission or dust continuum (Sect. \ref{subsec:cont_method}) are different lines detected at different redshifts. In this paper, we assume the lines detected by ALMA Band 8 are [\ion{C}{ii}] at similar redshifts to the central HzRGs.

\subsection{Continuum check}\label{subsec:cont_method}
We examine the continuum at the positions of each of the [\ion{O}{iii}] and [\ion{C}{ii}] companions (Table \ref{tab:summa_comp}). For potential companions, we may expect optical continuum emission, captured by NIRSpec IFU, which would presumably be dominated by evolved stellar populations. The ALMA Band 8 continuum (i.e., rest frame $\sim150\,\rm \mu m$) is emitted by cold dust which may trace forming stars. The case becomes more complicated with the presence of an AGN that could dominate the optical continuum \citep[][]{Wang_2024}. Excluding the companions near the radio AGN, we only find that J0121-[\ion{C}{ii}]-B shows optical continuum emission (Fig. \ref{fig:sp_0121_continuum_opt}). As for FIR continuum, we detect emission from J0121-[\ion{C}{ii}]-A, J0121-[\ion{C}{ii}]-B, and 4C19-[\ion{C}{ii}]. 

The most interesting case is from the field of \object{4C+03.24} (Sect. \ref{subsec:comp_4c03}). We discovered a tail-like continuum towards the south. It has a similar morphology to the systemic [\ion{O}{iii}]5007 emission. MUSE observation also captured UV continuum with a similar morphology (Appendix \ref{app:4C03_fore}).

\section{Results}\label{sec:results}
In this section, we report evidence for nearby companions around our radio AGN following the methods in Sect. \ref{sec:analysis}. Since the nature (i.e., interpretation of the gas dynamics) are unique for each target, we examine them individually. The properties of the companions detected in this work are summarized in Table \ref{tab:summa_comp}.

\subsection{Companions of TN J0121+1320}\label{subsec:comp_0121}
\object{TN J0121+1320} is the only target in our sample without a clearly resolved radio core. In this section, we first define the position of the radio AGN and then present evidence of nearby companions around it.

\subsubsection{Radio core of TN J0121+1320}\label{subsub:radiocore_0121}
The radio emission from \object{TN J0121+1320} has a compact morphology even in the highest resolution 8.4$\,$GHz dirty image ($\sim0.3\arcsec$). It only shows an elongation along the east-west direction (e.g., green contours in Fig. \ref{fig:O3_chanmap_0121}a\ref{fig:O3_chanmap_0121}b). We adopt the west peak of this elongation as the radio core based on: (i) its position being closer to the peak position of the line emission (e.g., [\ion{O}{iii}]$5007$); (ii) the morphology of [\ion{O}{iii}]$5007$ at $\Delta v=0\,\mathrm{km\,s^{-1}}$ resembling AGN ionization cone (Fig. \ref{fig:O3_chanmap_0121}a); (iii) its ALMA 100$\,$GHz continuum being skewed spatially towards the west \citep[][]{Falkendal_2019}. Argument (iii) is also based on the spectral energy distribution (SED) study of \citet{Falkendal_2019} where the ALMA Band 3 continuum was used to constrain the radio sychrotron emission and the cold dust continuum (see their Fig. A.7). Based on their analysis, the 100$\,$GHz Band 3 continuum may be contributed by both. If it indeed contains synchrotron emission, then the closer west 8.4$\,$GHz radio excess is likely to be the radio core of the radio AGN. Taking our new Band 8 continuum into consideration, the 100$\,$GHz continuum seems likely to trace more diffuse and colder dust. Even in this case, the host galaxy is more likely located towards the west where the dust is. 

\subsubsection{Results of companion search}

\citet{DeBreuck_2002} reported a $18-19\,$mag detection of \object{TN J0121+1320} in $K_{s}$ using Keck/NIRC, which may indicate the presence of nearby companions based on morphology (seeing $0.4\arcsec-0.5\arcsec$). We caution that the $K_{s}$ observation was likely contaminated by bright [\ion{O}{iii}]4959,5007 emission \footnote{\url{https://www2.keck.hawaii.edu/inst/nirc/manual/fw_macros.html}}. Based on a seeing-limited SINFONI $K-$band observation ($0.65\arcsec\times0.55\arcsec$), \citet{Nesvadba_2007a} reported two [\ion{O}{iii}]4959,5007 spatial components in the field of \object{TN J0121+1320} separated by $\sim10\,$kpc (component A and B under their naming). They also detected $K=19.8\pm$0.4$\,$mag continuum emission  at the emission peak of [\ion{O}{iii}]4959,5007 after excluding the line contamination (their component A). \citet{Nesvadba_2007a} suggested that component A may be the host galaxy of the radio AGN. We note that the continuum of radio AGN host galaxy may be contaminated by the foreground galaxy $\sim0.45\arcsec$ from the radio core (blue dot in Fig. \ref{fig:O3_chanmap_0121}a\ref{fig:O3_chanmap_0121}b, see also Appendix \ref{app:tnj0121_fore}). Though it may also be affected by leakage from MSA, we find that the continuum detected near the radio core in our NIRSpec IFU data is indeed spatially skewed towards the position of the foreground galaxy. 

We next focus on the new evidence of nearby companions in our latest datasets with sub-kiloparsec resolution that reveal that there may be three systems across the $\sim25\,$kpc FoV. To the west of the radio core of \object{TN J0121+1320}, we detect a $\sim$$6\,$kpc extended bright [\ion{O}{iii}] line-emitting nebula, J0121-[\ion{O}{iii}], (blue triangle in Fig. \ref{fig:O3_chanmap_0121}a\ref{fig:O3_chanmap_0121}b). As shown in the spectrum in Fig. \ref{fig:O3_chanmap_0121}c, the dominant emission component of J0121-[\ion{O}{iii}] is $434\,\rm km\,s^{-1}$ redshifted from the systemic velocity and has a width $\sigma_{v}=95\pm3\,\rm km\,s^{-1}$ (Table \ref{tab:1d_o3_spec_fit}). This is consistent with the results of \citet{Nesvadba_2007a} (i.e., their component B). J0121-[\ion{O}{iii}] is at a projected distance of $\sim12.6\,$kpc from the radio AGN, $d_{\rm RG}$, measured from the peak [\ion{O}{iii}] emission. Its narrower line width reveals its AGN-undisturbed nature. This suggests it is a distinct system from the gaseous medium surrounding the radio AGN and is likely a nearby companion. We map the kinematics of J0121-[\ion{O}{iii}] in Fig. \ref{fig:O3_chanmap_0121}d across $\sim25\,\rm kpc^{2}$, a clear velocity gradient along the north-south direction with $\Delta v\sim400\,\rm km\,s^{-1}$. Visually, the $\Delta v$ map does not resemble disk rotation in that. There is no clearly defined positive-negative velocity transition edge. $^{\rm 3D}$\textsc{Barolo} \citep[][]{DiTeodoro_2015} fitting of this [\ion{O}{iii}] emission nebula also does not indicate a clear rotation structure.

In addition to the [\ion{O}{iii}] companion, we detect three [\ion{C}{ii}] companions in this field (Fig. \ref{fig:j0121_c2_map_sp}). J0121-[\ion{C}{ii}]-A is at $d_{\rm RG}=17.2\,$kpc away from the radio core with a $\Delta v=-401\,\rm km\,s^{-1}$. Cold dust continuum is also detected at this position. We do not have information about warm ionized gas at this position since J0121-[\ion{C}{ii}]-A is outside the NIRSpec IFU FoV. J0121-[\ion{C}{ii}]-C is at $d_{\rm RG}=11.1\,$kpc and falls within the ionized nebula of J0121-[\ion{O}{iii}]. J0121-[\ion{C}{ii}]-C (Fig. \ref{fig:j0121_c2_map_sp}, Table \ref{tab:1d_c2_spec_fit}) has $\Delta v=366\,\rm km\,s^{-1}$ and a similar profile to the [\ion{O}{iii}]5007 line at the same position. Therefore,  J0121-[\ion{C}{ii}]-C and J0121-[\ion{O}{iii}] may trace the same companion system. J0121-[\ion{C}{ii}]-B has the brightest [\ion{C}{ii}] emission and is at $d_{\rm RG}=6.4\,$kpc. Compared to the [\ion{O}{iii}]5007 at the same position, J0121-[\ion{C}{ii}]-B may have different dynamics. 

To further study the cold ionized gas dynamics of J0121-[\ion{C}{ii}]-B, we performed spatial [\ion{C}{ii}] fitting using $^{\rm 3D}$\textsc{Barolo} \citep[][and Appendix \ref{app:barolo} for details]{DiTeodoro_2015}. The result is shown in Fig. \ref{fig:barolo_m1_0121} where we find a velocity gradient of $\sim 400\,\rm km\,s^{-1}$ along the north-south direction (i.e., PA$=187^{\circ}$ for the kinematic major axis). The residual from $^{\rm 3D}$\textsc{Barolo} and the position-velocity diagram  (especially moment 2 residual in Fig. \ref{fig:barolo_m02_j0121} and \ref{fig:barolo_pv_0121}) indicate gas motions other than disk rotation.

Both FIR and optical continuum are detected at J0121-[\ion{C}{ii}]-B (Fig. \ref{fig:O3_chanmap_0121} and \ref{fig:sp_0121_continuum_opt}). We show the aperture-extracted optical spectrum at J0121-[\ion{C}{ii}]-B. A stellar continuum fit is beyond the scope of this work. Instead, we overlay a model spectrum generated using Bagpipe \citep[][]{Carnall_2018_Bagpipe}. We used a single starburst model with solar metallicity. The extinction is set to $A_{V}=1.5\,$mag based on Balmer decrement and \citet{Calzetti_2000} law. We find that this continuum level could be consistent with a galaxy with $M_{\star}=10^{9.9}\,M_{\odot}$. This model spectrum should simply be considered as a zeroth order indication of the stellar mass.

\subsection{Companion search of TN J0205+2242}\label{subsec:comp_0205}

We detect a [\ion{O}{iii}] line-emitting blob, J0205-[\ion{O}{iii}], $d_{\rm RG}\sim3.22\,$kpc south of the radio AGN \object{TN J0205+2242} (Fig. \ref{fig:O3_chanmap_0205}a\ref{fig:O3_chanmap_0205}b). This kinematic feature was missed in previous SINFONI observations due to its limited spatial resolution and sensitivity \citep[$0.45\arcsec\times0.4\arcsec$ with adaptive optics correction,][]{Nesvadba_2007a}. J0205-[\ion{O}{iii}] has $\Delta v=-1426\,\rm km\,s^{-1}$ with respect to the systemic redshift of \object{TN J0205+2242}. It has a morphology with $\sim3\,$kpc extent which is barely resolved by the NIRSpec IFU. Compared to the spectrum of the central AGN, J0205-[\ion{O}{iii}] is a distinct emission component (cyan Gaussian, Fig. \ref{fig:O3_chanmap_0205}c). The fitted line width is relatively broad with $\sigma_{v}=309\,\rm km\,s^{-1}$ which could be affected by beam smearing. That is to say that the physical scale of J0205-[\ion{O}{iii}] cloud is small compared to the resolution of NIRSpec IFU, $\sim0.1-0.2\arcsec$ \citep[e.g.,][]{Vayner_2023Q3D,Vayner_2023Q3D_2,Vayner_2025}. As a result, the unresolved gas motion (e.g., velocity gradient due to rotation) artificially broadens the observed line width. J0205-[\ion{O}{iii}] has a tentative spatial velocity gradient along the northeast to southwest direction with $\sim300\,\rm km\,s^{-1}$ (Fig. \ref{fig:O3_chanmap_0205}d\ref{fig:O3_chanmap_0205}e). Another possibility is that this broad line width is due to the impact from the central AGN.

We detect tentative [\ion{C}{ii}] emission, J0205-[\ion{C}{ii}], $d_{\rm RG}=5.0\,$kpc north of the radio core at $3\sigma$ (Fig. \ref{fig:j0205_c2_map_sp}). The 1D spectrum shows a narrow line with $\sigma_{v}=54\,\rm km\,s^{-1}$ (Table \ref{tab:1d_c2_spec_fit}). Given low S/N, we take this as a marginal detection though there is a [\ion{O}{iii}] counterpart which may strengthens confidence in its existence (see the [\ion{O}{iii}] line profile in Fig. \ref{fig:j0205_c2_map_sp}b). The nature of this system would be more tentative if only one dataset were available. This showcases the advantage of joint ALMA-\textit{JWST} observation for discovering companion systems.

\subsection{Companions of 4C+03.24}\label{subsec:comp_4c03}

In this work, we use the new VLA 15$\,$GHz observation of \object{4C+03.24} to trace the jet morphology (Sect. \ref{subsec:radio_ima}). In this rest-frame 70$\,$GHz data, we resolve the central structures of jet where three spatial peaks are discovered \citep[Fig. \ref{fig:O3_chanmap_4c03}, see][for a comparison to 8$\,$GHz]{vanOjik_1996}. We note that a detailed radio analysis of \object{4C+03.24} is beyond the scope of this work and will be presented in Kukreti et al. (in prep.). We determine that the center one of these three peaks is the core, radio AGN, based on spectral slope (magenta star in Fig. \ref{fig:O3_chanmap_4c03}). The coordinate of the radio core is given in Table \ref{tab:sampleobs} which is used for WCS alignment (Sect. \ref{subsec:WCS_align}).

\object{4C+03.24} shows a complex warm ionized gas kinematics in the vicinity of its radio core. We focus on the most peculiar blob just to the north at $d_{\rm RG}=1.6\,$kpc but with a significant velocity shift, $\Delta v\simeq1300\,\rm km\,s^{-1}$ (precisely, $\Delta v=1395\pm47\,\rm km\,s^{-1}$ using fitting results from Table \ref{tab:1d_o3_spec_fit}). This emission nebula, 4C03-[\ion{O}{iii}], is barely resolved (Fig. \ref{fig:O3_chanmap_4c03}b). With \texttt{q3dfit}, we spectrally decompose it to a narrow and a broad component with $\sigma_{v}=55\pm16\,\rm km\,s^{-1}$ and $574\pm5\,\rm km\,s^{-1}$, respectively (Fig. \ref{fig:O3_chanmap_4c03}c lower panel and Table \ref{tab:1d_o3_spec_fit}). Due to low S/N, the spatial velocity fitting of the 4C03-[\ion{O}{iii}] narrow component is noisy, as shown in Fig. \ref{fig:O3_chanmap_4c03}d, but may show a tentative velocity gradient. The broad component of 4C03-[\ion{O}{iii}] shows a gradient along the southeast-northwest direction with $\Delta v\sim500\,\rm km\,s^{-1}$. Both the kinematics and morphology of the 4C03-[\ion{O}{iii}] broad component are distinct from the bulk motion of ionized gas detected by SINFONI \citep[][]{Nesvadba_2017b}, indicating a different nature from the jet-driven outflow. Taking both the narrow and broad components into account, we may be witnessing a companion AGN in the vicinity of the central radio quasar. Indeed, the 15$\,$GHz radio map shows a emission peak coincide with 4C03-[\ion{O}{iii}] northward of the radio core. This may further indicate that the presence of two radio AGN. Alternatively, this may suggest a strong jet gas interaction. 4C03-[\ion{O}{iii}] might be related to the previously published [\ion{C}{i}] emission of \object{4C+03.24} \citep[$\Delta v_{\rm [\ion{C}{i}]}=1123\pm 282\,\rm km\,s^{-1}$,][]{kolwa2023}.

We discovered three [\ion{C}{ii}] emitters despite the shallow ALMA observation. 4C03-[\ion{C}{ii}]-A is detected at $4\sigma$ and is located close to the radio core, $d_{\rm RG}=1.4\,$kpc. Since it is at the systemic redshift, 4C03-[\ion{C}{ii}]-A likely traces gas in the AGN host ($\Delta v=9\pm17\,\rm km\,s^{-1}$, Fig. \ref{fig:4c03_c2_map_sp}a\ref{fig:4c03_c2_map_sp}b). Base on the current dataset, we cannot exclude the possibility that 4C03-[\ion{C}{ii}]-A being a companion cloud system at projected distance near the core. 4C03-[\ion{C}{ii}]-B is at $d_{\rm RG}=15.8\,$kpc southeast of the radio core with $\Delta v=-99\,\rm km\,s^{-1}$. The [\ion{O}{iii}] profile at the same position has an emission component consistent with the [\ion{C}{ii}] emission. 4C03-[\ion{C}{ii}]-B is diffuse, as indicated by the clumpy contours shown in Fig. \ref{fig:4c03_c2_map_sp}a. 4C03-[\ion{C}{ii}]-C has an elongated morphology extended by $\sim10\,$kpc located to the southeast of the radio core and with $\Delta v=342\,\rm km\,s^{-1}$. The [\ion{O}{iii}] profile at the same position does not show a clear similarity. This may be a disturbed gas system in the vicinity of the radio AGN. We note that the rough positions of 4C03-[\ion{C}{ii}]-B and [\ion{C}{ii}]-C are given as the centroid positions of the $5\sigma$ contour (Table \ref{tab:summa_comp}).

 We show the optical continuum from NIRSpec IFU in Fig. \ref{fig:4c03_c2_map_sp}a as dashed green contours. This continuum map is constructed by collapsing the cube in wavelength directions with strong emission lines masked. We further masked artifacts at detector edges, the foreground galaxy to the west, and identified MSA leakage contamination.

The optical continuum has a complex morphology which resembles a merger tidal tail. A similar morphology is also seen in UV continuum by MUSE (Fig. \ref{fig:foreg_4c03} and also \textit{HST}). Taking 4C03-[\ion{C}{ii}]-B and -[\ion{C}{ii}]-C into account, it is more likely that we are witnessing stars and gas stripped through merger processes at the center of \object{4C+03.24}. 4C03-[\ion{O}{iii}] may also be in the same system. It has been known for decades that the southern radio jet of \object{4C+03.24} is bent \citep[e.g., Fig. \ref{fig:O3_chanmap_4c03}a, see also][]{vanOjik_1996}. The hypothesis is that a dense system may exist in the south deflecting the jet. With our new ALMA+\textit{JWST} data, we now confirm tidal structures to the south responsible for this deflection. A future paper will examine the jet-cloud interaction in further detail. The morphology of the extended continuum is similar to the [\ion{O}{iii}]5007 (Fig.\ref{fig:O3_chanmap_4c03})a. This cautions that the continuum emission might be scattered AGN light rather than stellar emission.

\subsection{Companion search of 4C+19.71}\label{subsec:comp_4c19}

\citet{Wang_2024} reported the detection of multiple emission components to the south of the radio core of \object{4C+19.71}, providing evidence for companions (see their Fig. 1c2). We confirm this detection at $d_{\rm RG}=6.2\,$kpc south of the core, 4C19-[\ion{O}{iii}] (Fig. \ref{fig:O3_chanmap_4c19}b). As seen in the aperture-extracted spectrum in Fig. \ref{fig:O3_chanmap_4c19}c, there are three line emission components. The most blueshifted component, shown as the teal Gaussian, has a spatial extent of only $\sim 3\,$kpc and is likely the companion. \citet{Wang_2024} suggested that the most redshifted component at the position of 4C19-[\ion{O}{iii}] as the receding emission from a rotating disk, with the most blueshifted one as the approaching side. After a spatial examination, we find that the two Gaussians redward of the teal Gaussian in Fig. \ref{fig:O3_chanmap_4c19}c are spatially more diffuse and are likely to be ionized ISM associated with the radio AGN. Hence, we infer that only the most blueshifted teal Gaussian is from a potential companion, with $\Delta v=-584\,\rm km\,s^{-1}$ and $\sigma_{v}=210\,\rm km\,s^{-1}$. In Fig. \ref{fig:O3_chanmap_4c19}d, we map the velocity shift 4C19-[\ion{O}{iii}]. We do not detect a clear gradient, though $\Delta v$ in the northern part of the blob shows evidence of being redshifted with respect to the southern part by $\sim100\,\rm km\,s^{-1}$ (Fig. \ref{fig:O3_chanmap_4c19}e). 

There is one bright [\ion{C}{ii}] emitter detected, 4C19-[\ion{C}{ii}] (Fig. \ref{fig:4c19_c2_map_sp}). FIR continuum peaking at the same position is also detected (Fig. \ref{fig:O3_chanmap_4c19}a\ref{fig:O3_chanmap_4c19}b). 4C19-[\ion{C}{ii}] is at a small projected distance from the radio core, $d_{\rm RG}=0.3\,$kpc ($0.04\arcsec$). We find that the [\ion{C}{ii}] and [\ion{O}{iii}] line profiles are different, as shown in Fig. \ref{fig:4c19_c2_map_sp}b (offset $\sim 300\,\rm km\,s^{-1}$). Based on fitting of the 1D spectrum, [\ion{C}{ii}] is at $\Delta v=-71\,\rm km\,s^{-1}$. This is more consistent with the redshift of the molecular gas ([\ion{C}{i}]) than the warm ionized gas and suggests that the cold and warm gas are associated with two different systems \citep[][]{Falkendal_2021,kolwa2023}. The absolute astrometric accuracy of ALMA can be estimated as $\mathrm{pos_{acc}} = \mathrm{beam_{FWHP}}/(\mathrm{S/N})/0.9$ where $\mathrm{beam_{FWHP}}$ is the Full Width Half Maximum synthesized beam size in arcseconds. With $\theta_{\rm 400GHz}=0.23\arcsec \times 0.18\arcsec$ and S/N$=13.3$, the positional accuracy is $\sim0.019\arcsec\simeq d_{\rm RG,4C19-[\ion{C}{ii}]}/2$. Assuming the errors are Gaussian distribution, the project distance is at $2\sigma$ away from the zero (i.e., the same position as the radio core) which implies that there is only a $5\%$ probability that the dust is at the same position as the AGN.  The probability can be even lower if we take the uncertainty of the radio position into consideration ($\sim \theta_{\rm radio}/10$). In addition, there is no information on the line-of-sight separation. Hence, 4C19-[\ion{C}{ii}] might be a gas-rich companion in front of or behind the massive radio galaxy \citep[e.g.,][]{Lin_2024}. We also model the [\ion{C}{ii}] emission with $^{\rm 3D}$\textsc{Barolo} and find evidence of disk rotation (kinematic major axis along east-west with PA=$91^{\circ}$, Fig. \ref{fig:barolo_m1_4c19} and Appendix \ref{app:barolo}). Based on the residuals, we cannot rule out the possibility that the [\ion{C}{ii}] kinematics is disturbed by a jet-driven cocoon if they indeed belong to the same system \citep[e.g.,][]{Mukherjee_2016}.

\begin{figure*}
    \centering
    \includegraphics[width=\textwidth,clip]{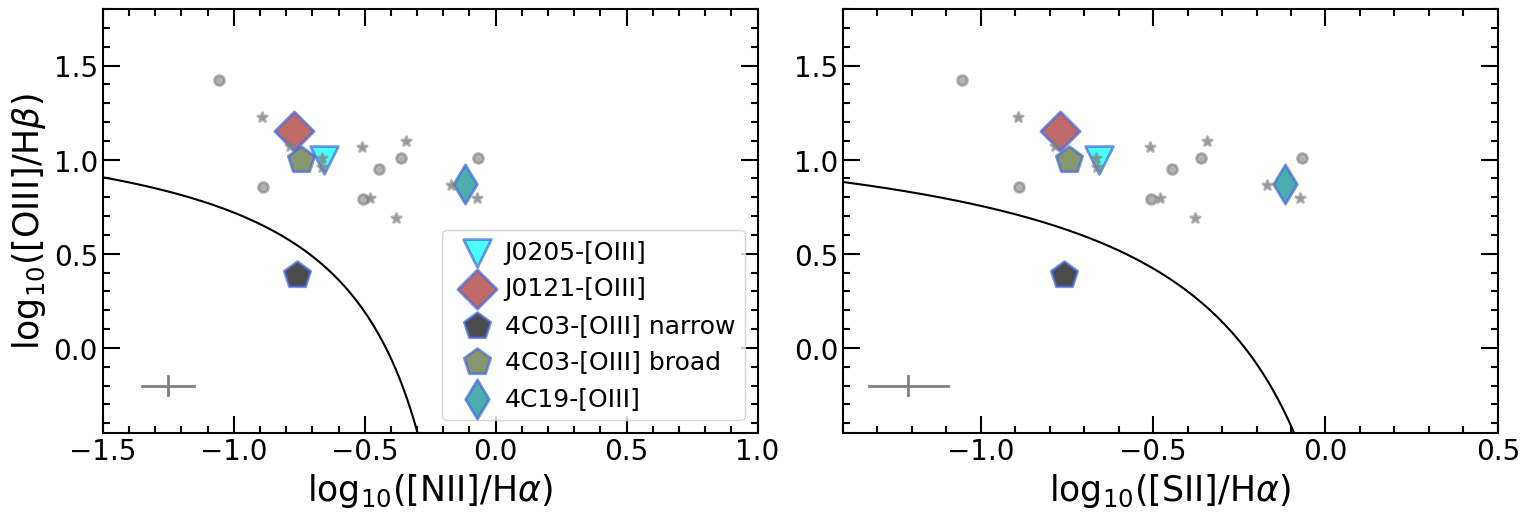}
    \caption{Line ratio diagnostic diagrams for the individual kinetic components from  \texttt{q3dfit} 1D spectral fitting at [\ion{O}{iii}] companion and radio core positions. We highlight the emission components of [\ion{O}{iii}] companions with larger symbols filled with same colors as used in Fig. \ref{fig:O3_chanmap_0121}c, \ref{fig:O3_chanmap_0205}c,, \ref{fig:O3_chanmap_4c03}c, and \ref{fig:O3_chanmap_4c19}c. The other kinetic components at radio core and companion positions are shown with smaller gray stars and dots, respectively. The black curves are the empirical classifications that separate quiescent galaxies (lower left) from AGN (above this line) \citep[][]{Kewley_2006}. Typical errors are shown in the bottom left corner.}\label{fig:BPT_1D}
\end{figure*}

\begin{table*}
 \caption{Properties of the companions detected in this work.}\label{tab:summa_comp}
 \centering
\begin{tabular}{l c c c c c c }
\hline
\hline

 \multicolumn{1}{c}{Companion} &  2nd detection &  Continuum  &  RA, Dec (J2000) &  $d_{\rm RG}$ & $M_{\rm dyn, dis}$ & $M_{\rm dyn, rot}$  \\
           &   & opt./FIR &  (hh:mm:ss, dd:mm:ss)  & arcsec/kpc       & $10^{10}\,M_{\odot}$                   & $10^{10}\,M_{\odot}$  \\
\multicolumn{1}{c}{(1)} & (2) & (3) & (4) & (5) & (6) & (7) \\
\hline
J0121- & & & & & & \\
\hline
[\ion{O}{iii}]  & \cmark  & \xmark/\xmark & (01:21:42.608, +13:20:58.52) & 1.72/12.6 & - & 3.1${\rm csc^{2}} i$  \\

[\ion{C}{ii}]-A & -       & -/\cmark      & (01:21:42.579, +13:20:59.26) & 2.36/17.2 & 4.5 & -  \\

[\ion{C}{ii}]-B & \xmark  & \cmark/\cmark & (01:21:42.667, +13:20:58.52) & 0.88/6.4  & 48 & 9.0  \\

[\ion{C}{ii}]-C & \cmark  & \xmark/\xmark & (01:21:42.622, +13:20:58.34) & 1.52/11.1 & 3.3  & -   \\
\hline
J0205- & & & & & & \\
\hline
[\ion{O}{iii}] &  \xmark & \xmark/\xmark  & (02:05:10.689, +22:42:50.17) & 0.44/3.2  & 22 &  0.8${\rm csc^{2}} i$  \\

[\ion{C}{ii}]  &  $\bigcirc$  & \xmark/\xmark  & (02:05:10.674, +22:42:51.26) & 0.69/5.0  & 1.8 & -  \\

\hline
4C03- & & & & & & \\
\hline
[\ion{O}{iii}] narrow & \xmark & \cmark/\xmark & (12:45:38.367, +03:23:21.45) & 0.22/1.6 & 0.7 & 0.3${\rm csc^{2}} i$   \\

[\ion{O}{iii}] broad & \xmark & \cmark/\xmark & - & - & 77  & 1.3${\rm csc^{2}} i$   \\

[\ion{C}{ii}]-A & \xmark & \xmark/\xmark & (12:45:38.377, +03:23:20.96)  & 0.20/1.4  & 4.3 & -   \\

[\ion{C}{ii}]-B & $\bigcirc$ & \xmark/\xmark & (12:45:38.470, +03:23:19.48)\tablefootmark{\rm{(\dag)}}   & 2.17/15.8 & 6.1 & -   \\

[\ion{C}{ii}]-C & \xmark & \xmark/\xmark & (12:45:38.420, +03:23:21.00)\tablefootmark{\rm{(\dag)}}   & 0.65/4.7  & 17 & -   \\
\hline
4C19- & & & & & & \\
\hline
[\ion{O}{iii}]  & \xmark & \xmark/\xmark & (21:44:07.515, +19:29:13.73)  & 0.85/6.17 & 10 & 0.2${\rm csc^{2}} i$   \\

[\ion{C}{ii}]   & \xmark & \cmark/\cmark & (21:44:07.515, +19:29:14.59)  & 0.04/0.3 & 32 & 7.0   \\
 \hline
\end{tabular}
\tablefoot{(1) Name of the companion. (2) Companion signature of the emission line other than the one used for detection. \cmark indicates the companion is detected by both methods. \xmark indicates there is no detection of the other line or there is no clear signature of companion from the other line.  $\bigcirc$ indicates that there is a hint of a companion from the other line. (3) Detection of optical or FIR continuum. (4) Companion coordinate. \tablefoottext{\rm{\dag}}{Position based on the $5\sigma$ contours, which should be treated with caution given the elongated (disrupted) morphology.} (5) Projected distance between the radio core and the position of the companion. (6) Dynamical mass of the companion based on velocity dispersion. (7) Dynamical mass of the companion assuming disk rotation.
}
\end{table*}

\section{Discussion} \label{sec:discussion}

\subsection{Companion kinematics and masses}\label{subsec:comp_mass}

Companions play an important role in galaxy evolution. In particular, nearby companion systems as detected in this work may be mergers which are key for both mass growth and triggering AGN \citep[][]{Springel_2005, Hopkins_2008}. The mass of a companion system is a crucial parameter. With our new observations, we offer a first order mass estimate. For this calculation, we ignore gravitational interactions between the companion and the central AGN.

Most of the companions have large velocity dispersion, $\sigma_{v}$, which may indicate they are in dispersion dominated systems \citep[$v_{\rm rot}/\sigma_{v}\lesssim\sqrt{3.36}$, where $v_{\rm rot}$ is the rotation velocity, e.g.,][and see the fitting in Appendix \ref{app:1d_linefit}]{Forster-Schreiber_2020}. Under this assumption, we estimate $M_{\rm dyn}$ following \cite{ubler_2023}: 
\begin{equation}\label{eq:m_dyndis}
    M_{\rm dyn, dis} = K(n)K(q)\frac{\sigma_{v}^{2}r_{e}}{G}
\end{equation}
where $G$ is the gravitational constant, $r_{e}$ is the effective radius, $K(n)=8.87-0.831n+0.0241n^{2}$ is a function of S\'{e}rsic index, $n$, and $K(q)=[0.87+0.38e^{-3.71(1-q)}]^{2}$ is a function of axis ratio, $q$. For unresolved systems, we use the PSF or major synthesized beam size (or 2$\theta$ for J0205-[\ion{C}{ii}] and 4C03-[\ion{C}{ii}]) as a proxy for $r_{e}$. We stress that this mass estimate should be treated as an upper limit, as the size is overestimated. For S\'{e}rsic index we use $n=1$ which gives an upper limit of $M_{\rm dyn}$. For simplicity, we take $q=1$. Equation \ref{eq:m_dyndis} implies $M_{\rm dyn, dis}$ for the companions discovered in this work are approximately $10^{10-11}\,M_{\odot}$ (Table \ref{tab:summa_comp}).

Alternatively, we can calculate the dynamical mass of the companion under the assumption of disk rotation using 
\begin{equation}
    M_{\rm dyn, rot} = k\frac{r v_{\rm rot}(r)^{2}}{G}
\end{equation}
where $v_{\rm rot}(r)$ is the circular rotation velocity, and $k$, the virial coefficient, is a function of $n$ and $q$. We take $k=1.8$ \citep[e.g.,][]{deGraaff_2024}. For J0121-[\ion{C}{ii}]-B and 4C19-[\ion{C}{ii}], we obtain $r$ and  $v_{\rm rot}$ from $^{\rm 3D}$\textsc{Barolo} where the disk inclination angle, $i$, has been taken into account. $M_{\rm dyn}$ can be directly calculated for these two (Sect. \ref{subsec:comp_0121} and \ref{subsec:comp_4c19}). For the other fainter [\ion{C}{ii}] companions, we do not estimate a mass under the rotation scenario due to limited information. Warm ionized gas was also used to study regular dynamics \citep[][]{Ubler_2019,deGraaff_2024}. For the [\ion{O}{iii}] companions, we found tentative velocity gradients from \texttt{q3dfit} (Sect. \ref{sec:results}). Assuming that these are rotating disks, we can perform the similar $M_{\rm dyn}$ estimation. Estimation of $v_{\rm rot}$ for these [\ion{O}{iii}] companions is nontrivial  given their marginally resolved spatial extents (except J0121-[\ion{O}{iii}]). Another uncertainty comes from their unknown inclination, $i$. Without further information, we use $v_{\rm rot}=\frac{1}{2}\Delta v_{\rm proj}/{\rm sin}\,i$ as a proxy for the rotation velocity where $\Delta v_{\rm proj}$ is the projected velocity difference along the gradient and leaving ${\rm sin}\,i$ unconstrained \citep[][]{Nesvadba_2011}. Hence, the equation we use is  
\begin{equation}
    M_{\rm dyn, rot} = k\frac{\Delta v_{\rm proj}^{2}R}{4G}{\rm csc^{2}}\,i
\end{equation}
where ${\rm csc^{2}}\,i=1/{\rm sin^{2}}\,i$ is left unconstrained. For the unresolved systems, we take the size of the NIRSpec IFU PSF, $\sim0.11\arcsec$, as the disk size, $R$, which is a lower limit. This simple estimation implies $M_{\rm dyn, rot}$ is around $10^{9-10}\,M_{\odot}$ (Table \ref{tab:summa_comp})

As shown in Sect. \ref{subsec:comp_0121} and Fig. \ref{fig:sp_0121_continuum_opt}, we estimate $M_{\star}$ estimation of J0121-[\ion{C}{ii}]-B to be $10^{9.9}\,M_{\odot}$. This is the only companion where a zeroth order estimation could be done based on the continuum detected by NIRSpec IFU. By integrating along the wavelength direction of the NIRSpec IFU cube excluding strong emission lines, we find that the continuum of J0121-[\ion{C}{ii}]-B is detected at $\sim8\sigma$. This indicates that the $3\sigma$ $M_{\star}$ detection limit in the cube is $\sim10^{9.5}\,M_{\odot}$ if optical continuum flux is directly proportional to stellar mass. Since our four radio AGN systems are at similar redshifts and observed with similar exposure times, this limit should be applicable to all four fields, with the caveat that the dust extinction of J0121-[\ion{C}{ii}]-B is high, implying the $M_{\star}$ detection upper limit may be lower if extinction is lower in the other fields. For J0121-[\ion{C}{ii}]-B, its $M_{\rm dyn}$ is around $10^{11}\,M_{\odot}$ which gives a $M_{\star}/M_{\rm dyn}\sim0.1$. At high-$z$, we may find that galaxies with low $M_{\star}$ have high $M_{\rm gas}/M_{\rm dyn}$ \citep[][]{Pillepich_2019, deGraaff_2024}. Hence, the other companion systems may have low stellar mass ($\lesssim10^{9.5}\,M_{\odot}$) and/or may be extremely gas-rich (or pure gas cloud systems).

Despite the large uncertainties in mass estimates, we infer that the companions discovered here are a few percent to the central host galaxies of the radio AGN \citep[$M_{\star}\gtrsim10^{11}\,M_{\odot}$][]{Seymour_2007,DeBreuck_2010,Falkendal_2019,Wang_2021}. If they will go through mergers, these will be the minor mergers which are critical for the hierarchal growth of clusters \citep[][]{Moster_2018}. Despite previous environmental studies on cluster scales around larger sample of HzRGs which proves their proto-cluster natures \citep[e.g.,][]{Venemans_2007,Wylezalek_2013,Wylezalek_2014}, there is no clear evidence that the four HzRGs studied here live in dense environments. Our observations bring new information on 5 to $20\,$kpc scales and suggest the existence of $M_{\rm dyn}\sim10^{10}\,M_{\odot}$ systems around these four radio galaxies. Given that the detection presented in this work is not complete, the number of companion systems could be higher. If these companions will merge with the central mass galaxies, we are observing the forming progenitors of $\gtrsim10^{12-13}\,M_{\odot}$ halos, traced by radio AGN beyond Cosmic Noon \citep[an estimate using $M_{\star}/M_{\rm halo}\sim 0.01-0.1$, e.g.,][]{Behroozi_2010,Behroozi_2019}. The detection rate of companions in our randomly selected sample is $100\%$. As a rough first order estimate, the major merger fraction in our four-source sample is $f_{\rm MM}\sim25\%$, since 4C03 may be a potential major merger.

The discussion above ignores the gravitational impact from the central AGN. At the physical scales probed in this paper, it is very likely that these companion systems are disturbed in the potential well. Supporting this, the residuals from $^{\rm 3D}$\textsc{Barolo} fitting of J0121-[\ion{C}{ii}]-B and 4C19-[\ion{C}{ii}] indicate the existence of complex gas motions.

\subsection{Nature of [\ion{O}{iii}] companions}\label{subsec:discu_agn}

Unlike [\ion{C}{ii}] companions indicating cold gas systems, the nature of [\ion{O}{iii}] companions is often elusive in the vicinity of luminous quasars with $L_{\rm bol}\gtrsim10^{47}\,\rm erg\,s^{-1}$ \citep[][]{Falkendal_2019}. [\ion{O}{iii}] emission with complex features are discovered in bright $z\sim0.5$ type-2 AGN, which has been interpreted as debris from mergers \citep[e.g., $L_{\rm bol}\sim^{46}\,\rm erg\,s^{-1}$,][]{Storchi-Bergmann_2018,Dallagnol_2021}. Our detection criteria for [\ion{O}{iii}] companions requires peculiar kinematics distinct from the bulk motion of the warm ionized gas \citep[][]{Nesvadba_2017b}. As a first order check of the ionizing mechanism, we investigate line ratio diagnostics based on our 1D spectral fitting \citep[Sect. \ref{sec:analysis},][]{Kewley_2001,Kewley_2006}. We use both the [\ion{O}{iii}]/H$\beta$-[\ion{N}{ii}]/H$\alpha$ and [\ion{O}{iii}]/H$\beta$-[\ion{S}{ii}]/H$\alpha$ diagnostics. The line fluxes are corrected for extinction using Balmer decrement by adopting the canonical value of $F_{\rm H\alpha}/F_{\rm H\beta}=3.1$ assuming Case B recombination \citep[][]{Veilleux_1987,Osterbrock_2006}. The results are shown in Fig. \ref{fig:BPT_1D}. Simply assuming the \citet{Kewley_2006} empirical classification, we conclude that all but one [\ion{O}{iii}] companion are photonionized by an AGN. Within the uncertainties of the line fitting and classification, the narrow component of 4C03-[\ion{O}{iii}] is also consistent with AGN photoionization. One immediate conclusion is that these emission components are indeed in the vicinity of powerful AGN, particularly J0205-[\ion{O}{iii}] and 4C03-[\ion{O}{iii}] with $|\Delta v|\sim1000\,\rm km\,s^{-1}$. 

Except for J0121-[\ion{O}{iii}] whose morphology, kinematics, and [\ion{C}{ii}] detection make it a solid companion detection, we cannot rule out that the candidates are part of the outflows. Though they are all detected along the jet axis, the PA of tentative velocity gradients are not always clearly correlated with the jet axis which may support them being companions. The relatively large $|\Delta v|\sim1000\,\rm km\,s^{-1}$ between these companions and the radio AGN may be expected, which is consistent with the velocity range of companion motions in clusters \citep[e.g.,][]{Bahcall_1994}. Hence, a possible scenario could be that there are more of these gaseous companion systems moving randomly in the potential well of the proto-cluster centered around the radio AGN. 
In this scenario, the detected companions that happen to fall in the ionization cone are photoionized and therefore detected \citep[i.e., the jet axis traces the direction of the ionization cone,][]{Drouart_2012}. Another scenario could be that the [\ion{O}{iii}] blobs originated from gas stripped from companions in a previous (or ongoing) interaction, and is now being illuminated by the AGN (e.g., 4C03-[\ion{O}{iii}]).

We do not intend to be complete in companion detection in this paper but offer a possible explanation of these unresolved emission blobs with peculiar kinematics. We acknowledge that the [\ion{O}{iii}] companions are clumps accelerated or illuminated by the AGN but may not in themselves be stellar systems (Sect. \ref{sec:analysis}).

\subsection{Impact on galaxy evolution}\label{subsec:dis_merger}

There is no clear agreement on how powerful AGN jets are triggered in the early universe. Merger processes provide one possibility \citep[e.g.,][]{Chiaberge_2015}, though the confirmation of mergers and/or interactions is nontrivial without 3D information. In this work, we discover that the companions have a median $d_{\rm RG}\simeq7\,$kpc. Using a typical $\Delta v$ of $\sim100-500\,\rm km\,s^{-1}$ and assuming no projection, the zeroth order estimate of encounter time is around 10 to 100$\,$Myr (i.e., $d_{\rm RG}/\Delta v$). Assuming the companions are in the post-encounter phase, i.e., after the first pericenter passage, this time scale is consistent with the observed time after jet launch \citep[e.g.,][]{Nesvadba_2017b,Nesvadba_2017a}. This supports the merger trigger scenario for jets.

Our detections show that there are gaseous clouds at $\lesssim18\,$kpc from these four HzRGs. With resolution limited observations, \citet{Falkendal_2019} and \citet{kolwa2023} reported that gas in radio AGN hosts should be relatively depleted. Taking this into account, our work may support the scenario that gas, if not completely depleted, is located in nearby systems rather than in the AGN hosts. This is consistent with literature studies of high-$z$ radio AGN, which show that gas-rich systems are present in the vicinity of massive host galaxies \citep[e.g.,][]{Emonts_2015,Lin_2024,Huang_2024}. Our work may show evidence of how these mergers with high gas content evolve in the radiation field of AGN and contribute to the growth of mass of the system. \citet{Lehnert_2016} predicted that $z\sim2-4$ galaxies are gas-rich (i.e., with \ion{H}{i}). These galaxies with $M_{\star}\sim10^{8-9}\,M_{\odot}$ are important for mass growth through minor merger processes. That is to say, gas-rich mergers like the ones discovered in this paper bring fuel to feed central SMBHs and lead to AGN ignition. Besides bringing gas within tens of kiloparsec scales, the disruption from these nearby companions plays an important role in dissipating angular moment and accelerating gas accretion.

Apart from triggering AGN activity, companions could alter jet development. \object{TN J0121+1320} is the only HzRG in our sample with a compact jet (at $0.3\arcsec$ resolution). All three of its companions are to the west of the radio source. Dust is also detected at the same location as the dense environment. Hence, we infer that interaction of the jet with the dense medium disturbs the jet propagation, making the jet flow more turbulent and the emission more diffuse, thus also reducing its surface brightness. Furthermore, the dense environment may even cause the jet of \object{TN J0121+1320} to fail in penetrating through the medium, resulting in its observed compact morphology despite being one of the most luminous radio sources \citep[$L_{\rm 500\,MHz}=28.49\,\rm W\,Hz^{-1}$,][]{DeBreuck_2010}. The case of \object{4C+03.24} may be a prototypical example. It has gone through a merger phase during which gas (4C03-[\ion{C}{ii}]-B and -[\ion{C}{ii}]-C) and stars (continuum) were stripped. This merger process triggered the radio jet which is then deflected by debris to the south.

We consider how radio selection effects may cause companions to be identified in particular phases. \citet{Emonts_2023_CO} examined the correlation between radio axis and gas clouds or mergers and showed that the radio asymmetries (e.g., radio flux ratios) can be attributable to interactions with gas in the CGM. Our work on smaller scales similarly suggests that the gaseous systems deflect or disturb jets, resulting in asymmetry, for example, by brightening radio fluxes. This means that flux-limited radio surveys will preferentially detect radio galaxies that are interacting because this resistance causes the synchrotron emissivity to increase by driving shocks into the relativistic plasma \citep[increasing the density of field lines and of radiating particles, see also, e.g.,][]{Steffen_1997,villar-martin_1999,Best_1998,Gomez_2006,Anderson_2022_spradio,Jerrim_2024}.

\section{Conclusion}\label{sec:conclusion}
We present the first analysis of the resolution-matched joint \textit{JWST}/NIRSpec IFU and ALMA observations of a sample of four HzRGs at $z\approx3.5$. We focus on [\ion{O}{iii}]5007 and [\ion{C}{ii}] emitters detected by the two observatories (Sect. \ref{sec:analysis}) and report the ubiquitous discovery of companion systems within $\sim18\,$kpc of the powerful radio AGN using these two independent methods.  

Specifically, we examine the 1D spectra at the positions of companions and quantify the gas kinematics ($\Delta v$ and $\sigma_{v}$). For [\ion{O}{iii}] companions (Sect. \ref{subsec:O3_kine}), we map the gas kinematics with \texttt{q3dfit}. We discover velocity gradients showing different motions than the bulk of the ionized gas. For the two [\ion{C}{ii}] companions detected at high S/N (J0121-[\ion{C}{ii}]-B and 4C19-[\ion{C}{ii}] with S/N$>$10), we perform $^{\rm 3D}$\textsc{Barolo} fits and detect disk rotation with residuals suggesting more chaotic underlying motions (Sect. \ref{subsec:c2_detection}). 

We discuss individual companions in Sect. \ref{sec:results}. First order dynamical mass estimates derived from the gas kinematics imply a range of $10^{9-11}\,M_{\odot}$, suggesting a possible minor merger with the central radio AGN host galaxy (Sect. \ref{subsec:comp_mass}). We further discuss the nature of [\ion{O}{iii}] companions in Sect. \ref{subsec:discu_agn} which indicate AGN photoionized gaseous clouds that happen to be within the ionization cone, though we cannot rule out the possibility that they are part of an AGN outflow. Finally, we discuss that the companions could trigger the launch of a jet by supplying gas and dissipating angular moment. These companions systems can also alter the jet development. This further suggests that these relatively minor companions play an important role in the evolution of massive systems.

The matched resolution \textit{JWST} and ALMA data are very rich and more detailed continuum and emission line studies are in preparation to decipher these complex systems. This is critical to understand the AGN feedback, star formation quenching, and mass growth of massive systems in the early universe.

\begin{acknowledgements}
We thank the anonymous referee for their valuable comments and suggestions, which have improved the quality of this manuscript. 

This work uses the NASA's Astrophysics Data System and a number of open source software other than the aforementioned ones such as Jupyter notebook \citep[][]{kluyver2016jupyter}; \texttt{matplotlib} \citep[][]{Hunter_2007}; \texttt{SciPy} \citep[][]{virtanen2020scipy}; \texttt{NumPy} \citep[][]{harris2020numpy}; \texttt{Astropy} \citep[][]{Astropy_2018}; \texttt{LMFIT} \citep[][]{newville2016}. We thank Hannah {\"U}bler for the discussion of ionized gas kinematics. We thank Yechi Zhang for the discussion of modeling optical stellar continuum.

This publication has received funding from the European Union’s Horizon 2020 research and innovation programme under grant agreement No 101004719 (ORP).

This work is based in part on observations made with the NASA/ESA/CSA James Webb Space Telescope. The data were obtained from the Mikulski Archive for Space Telescopes at the Space Telescope Science Institute, which is operated by the Association of Universities for Research in Astronomy, Inc., under NASA contract NAS 5-03127 for JWST. These observations are associated with program JWST-GO-01970. Support for program JWST-GO-01970 was provided by NASA through a grant from the Space Telescope Science Institute, which is operated by the Association of Universities for Research in Astronomy, Inc., under NASA contract NAS 5-03127.

D.W. and W.W. acknowledge support through an Emmy Noether Grant of the German Research Foundation, a stipend by the Daimler and Benz Foundation and the MERAC foundation and a Verbundforschung grant by the German Space Agency.  

W.W. also acknowledges the grant support from NASA through JWST-GO-3045 and JWST-GO-3950.

The work of D.S. was carried out at the Jet Propulsion Laboratory, California Institute of Technology, under a contract with NASA. 

B. D. O. acknowledges the support from the Coordena{\c c}{\~a}o de Aperfei{\c c}oamento de Pessoal de N\'ivel Superior (CAPES-Brasil, 88887.985730/2024-00).

This paper makes use of the following ALMA data: ADS/JAO.ALMA\#2021.1.00576.S. ALMA is a partnership of ESO (representing its member states), NSF (USA) and NINS (Japan), together with NRC (Canada), MOST and ASIAA (Taiwan), and KASI (Republic of Korea), in cooperation with the Republic of Chile. The Joint ALMA Observatory is operated by ESO, AUI/NRAO and NAOJ.
\end{acknowledgements}

\bibliographystyle{aa}
\bibliography{references}

\begin{appendix}
\section{[\ion{O}{iii}]5007 channel maps}\label{app:o3_channel}
\begin{figure*}
    \centering
    \includegraphics[width=\textwidth,clip]{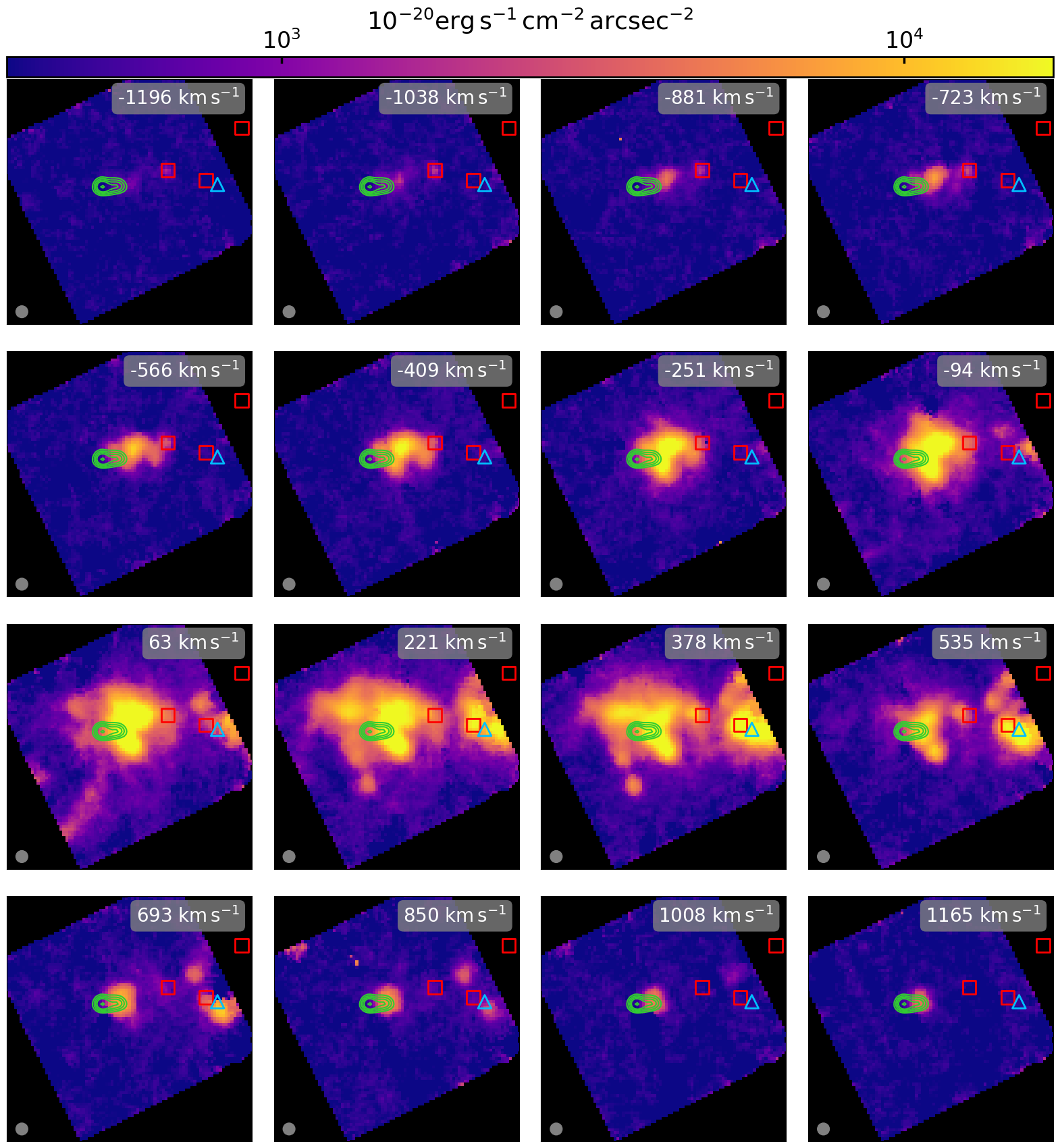}
    \caption{[\ion{O}{iii}]5007 channel maps for \object{TN J0121+1320}. We overlay the radio contours in green. The positions of [\ion{O}{iii}] (blue triangle) and [\ion{C}{ii}] (red square) companions are shown. }\label{fig:o3_stamps_0121}
\end{figure*}

\begin{figure*}
    \centering
    \includegraphics[width=\textwidth,clip]{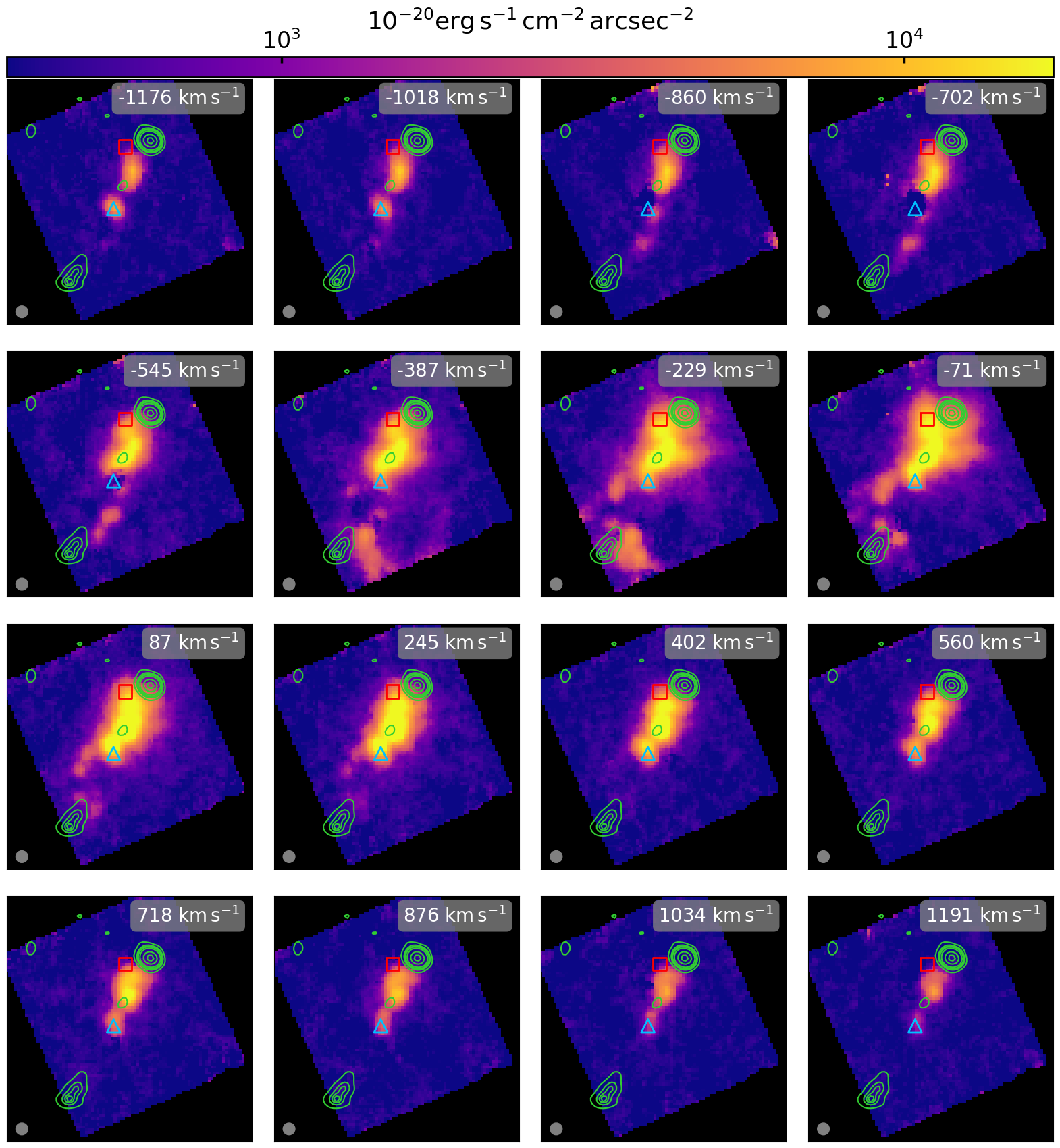}
    \caption{Similar to Fig. \ref{fig:o3_stamps_0121}, [\ion{O}{iii}]5007 channel maps for \object{TN J0205+2242}. We note that there are some bad pixels in channels $-860$ and $-702\,\rm km\,s^{-1}$. }\label{fig:o3_stamps_0205}
\end{figure*}

\begin{figure*}
    \centering
    \includegraphics[width=\textwidth,clip]{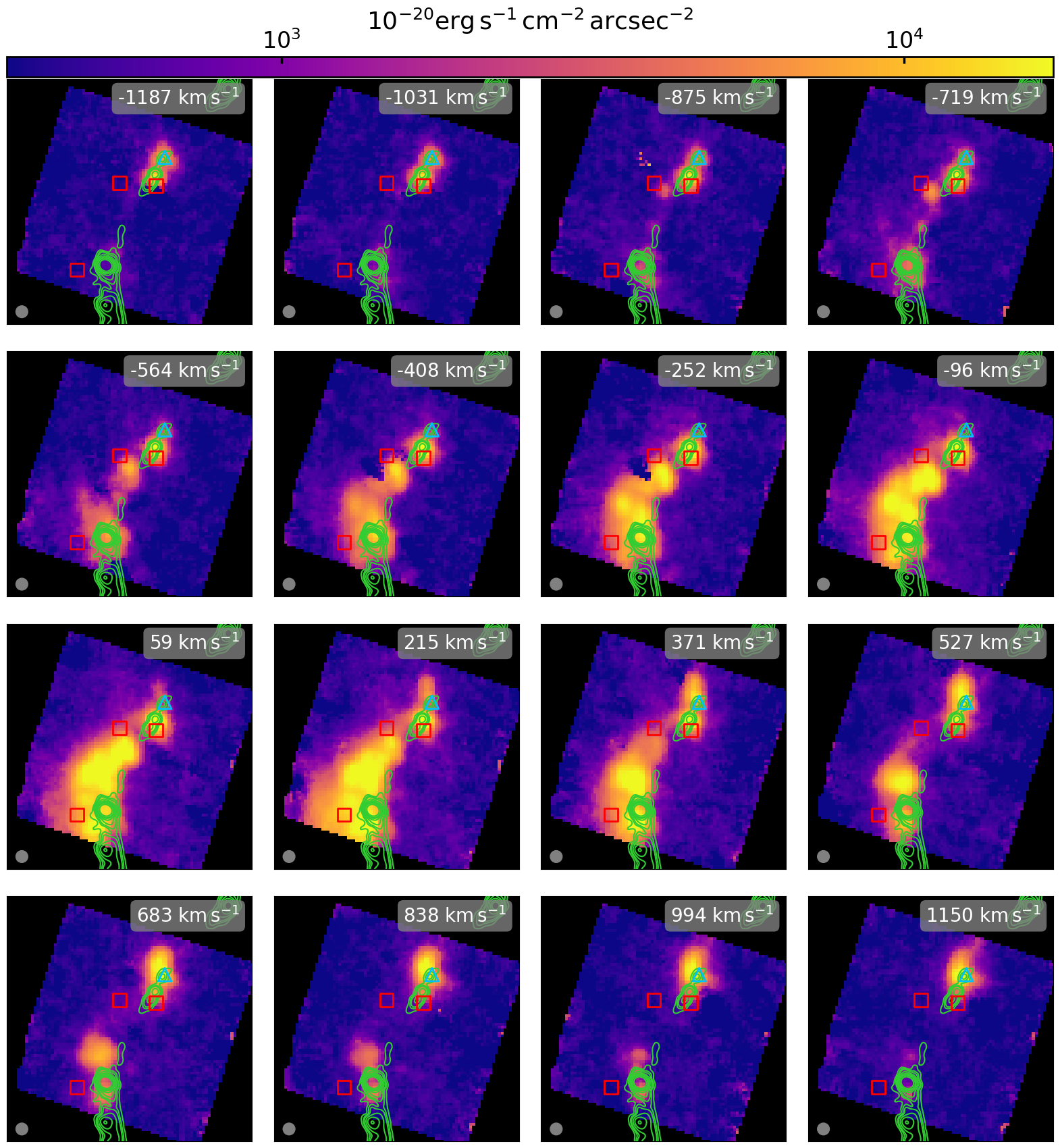}
    \caption{Similar to Fig. \ref{fig:o3_stamps_0121}, [\ion{O}{iii}]5007 channel maps for \object{4C+03.24}.}\label{fig:o3_stamps_4C03}
\end{figure*}

\begin{figure*}
    \centering
    \includegraphics[width=\textwidth,clip]{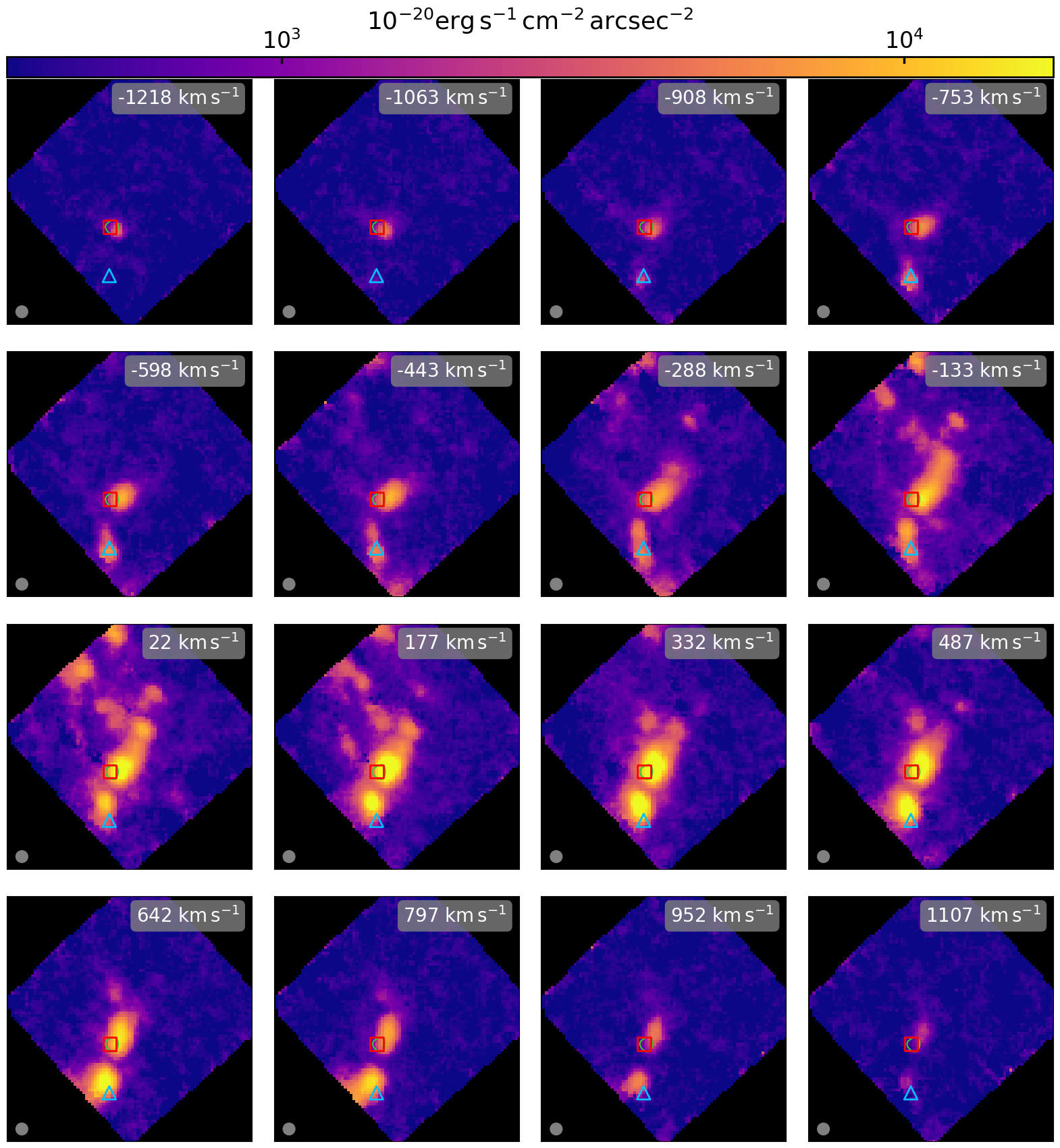}
    \caption{Similar to Fig. \ref{fig:o3_stamps_0121}, [\ion{O}{iii}]5007 channel maps for \object{4C+19.71}. }\label{fig:o3_stamps_4C19}
\end{figure*}

We present the [\ion{O}{iii}]5007 channel maps in this section. The channel maps are constructed from continuum subtracted NIRSpec IFU data cubes. Each map is collapsed from three consecutive channels. The maps shown in Fig. \ref{fig:o3_stamps_0121}, \ref{fig:o3_stamps_0205}, \ref{fig:o3_stamps_4C03}, and \ref{fig:o3_stamps_4C19} are for \object{TN J0121+1320}, \object{TN J0205+2242}, \object{4C+03.24}, and \object{4C+19.71}, respectively.

\section{1D emission line fitting results}\label{app:1d_linefit}
The fitted parameters of the 1D spectra are given in this section. For the warm ionized gas spectra extracted from the NIRSpec IFU, we present the redshifts and line widths for each Gaussian component in Table \ref{tab:1d_o3_spec_fit}. The fitted results of 1D [\ion{C}{ii}] spectra are presented in Table \ref{tab:1d_c2_spec_fit}.

\begin{table}
 \caption{Ionized gas kinematics from the 1D spectra extracted at the positions of the radio core and [\ion{O}{iii}] companions.}\label{tab:1d_o3_spec_fit}
 \centering
\begin{tabular}{c c c l }
\hline
\hline

 component  & redshift & \multicolumn{1}{c}{$\Delta v$} & \multicolumn{1}{c}{$\sigma_{v}$}  \\
            & $z$        & \multicolumn{1}{c}{$\rm km\,s^{-1}$} & \multicolumn{1}{c}{$\rm km\,s^{-1}$} \\
\hline
\hline
\multicolumn{4}{c}{TN J0121-radio core} \\
\hline
0  & 3.5191$\pm$0.0003 & 9$\pm$17   &191$\pm$11  \\
1  & 3.5243$\pm$0.0002 & 352$\pm$10 &132$\pm$7   \\
2  & 3.5246$\pm$0.0002 & 373$\pm$12 &628$\pm$10   \\
\hline
\hline
\multicolumn{4}{c}{J0121-[\ion{O}{iii}]} \\
\hline
0  & 3.5236$\pm$0.0003 & 305$\pm$20 & 248$\pm$13  \\
1\tablefootmark{\rm{(\dag)}}  & 3.5255$\pm$0.0001 & 434$\pm$2 &95$\pm$3   \\
\hline
\hline
\multicolumn{4}{c}{TN J0205-radio core} \\
\hline
0   & 3.5015$\pm$0.0001 & -302$\pm$4 & 158$\pm$5  \\
1   & 3.5059$\pm$0.0001 & -4$\pm$7   & 508$\pm$9  \\
2   & 3.5091$\pm$0.0001 & 204$\pm$2  & 106$\pm$2  \\
\hline
\hline
\multicolumn{4}{c}{J0205-[\ion{O}{iii}]} \\
\hline
0\tablefootmark{\rm{(\dag)}}   & 3.4846$\pm$0.0001 & -1426$\pm$8 & 309$\pm$7  \\
1  & 3.5066$\pm$0.0003 & 42$\pm$18 &618$\pm$21  \\
2  & 3.5085$\pm$0.0001 & 167$\pm$4 &174$\pm$6  \\
\hline
\hline
\multicolumn{4}{c}{4C03-radio core} \\
\hline
0  & 3.5533$\pm$0.0002 & -817$\pm$16 &1092$\pm$12  \\
1  & 3.5639$\pm$0.0007 & -115$\pm$46 &294$\pm$21  \\
2  & 3.5691$\pm$0.0002 & 222$\pm$15  &204$\pm$14  \\
\hline
\hline
\multicolumn{4}{c}{4C03-[\ion{O}{iii}]} \\
\hline
0  & 3.5748$\pm$0.0001 & 595$\pm$4 &158$\pm$6  \\
1\tablefootmark{\rm{(\dag)}} & 3.5852$\pm$0.0002 & 1280$\pm$11 &55$\pm$16  \\
2\tablefootmark{\rm{(\dag)}} & 3.5872$\pm$0.0001 & 1410$\pm$9  &574$\pm$5  \\
\hline
\hline
\multicolumn{4}{c}{4C19-radio core} \\
\hline
0  & 3.5822$\pm$0.0021 & -455$\pm$134 &712$\pm$54  \\
1  & 3.5933$\pm$0.0001 & 265$\pm$3    &207$\pm$4     \\
2  & 3.5947$\pm$0.0012 & 361$\pm$75   &444$\pm$49  \\
\hline
\hline
\multicolumn{4}{c}{4C19-[\ion{O}{iii}]} \\
\hline
0\tablefootmark{\rm{(\dag)}}  & 3.5803$\pm$0.0001 & -584$\pm$9 &210$\pm$9  \\
1  & 3.5874$\pm$0.0001 & -121$\pm$7 &98$\pm$8     \\
2  & 3.5970$\pm$0.0001 & 512$\pm$3  &213$\pm$3  \\

\hline
\end{tabular}
\tablefoot{Kinematic components of each sources are ordered based on their redshifts. $\Delta v$ is calculated with respect to the systemic redshift listed in Table \ref{tab:sampleobs}. The quoted uncertainties are $1\sigma$ errors from \texttt{q3dfit}. 
\tablefoottext{\rm{\dag}}{Kinematic emission from of the potential [\ion{O}{iii}] companion. } \\
}
\end{table}

\begin{table*}
 \caption{Fitting results of 1D [\ion{C}{ii}] spectra.}\label{tab:1d_c2_spec_fit}
 \centering
\begin{tabular}{c c c c c}
\hline
\hline

 Companion  & redshift & \multicolumn{1}{c}{$\Delta v$} & \multicolumn{1}{c}{$\sigma_{v}$} & $F_{\nu \rm [\ion{C}{ii}]}\Delta v$ \\
            & $z$        & \multicolumn{1}{c}{$\rm km\,s^{-1}$} & \multicolumn{1}{c}{$\rm km\,s^{-1}$} & $\rm Jy\, km\,s^{-1}$\\
\hline
\hline
J0121-[\ion{C}{ii}]-A   & 3.5130$\pm$0.0003 & -401$\pm$17 & 105$\pm$17 & 0.49$\pm$0.07 \\
J0121-[\ion{C}{ii}]-B   & 3.5197$\pm$0.0003 & 49$\pm$19 & 345$\pm$19 & 2.49$\pm$0.12 \\
J0121-[\ion{C}{ii}]-C   & 3.5245$\pm$0.0003 & 366$\pm$18 & 90$\pm$18 & 0.33$\pm$0.06\\
\hline
J0205-[\ion{C}{ii}]   & 3.5053$\pm$0.0003 & -44$\pm$19 & 54$\pm$19 & 1.11$\pm$0.34 \\
\hline
4C03-[\ion{C}{ii}]-A   & 3.5658$\pm$0.0003 & 9$\pm$17 & 66$\pm$17 & 0.65$\pm$0.14\\
4C03-[\ion{C}{ii}]-B   & 3.5690$\pm$0.0003 & 219$\pm$18 & 73$\pm$18 & 0.68$\pm$0.14\\
4C03-[\ion{C}{ii}]-C   & 3.5709$\pm$0.0003 & 343$\pm$21 & 117$\pm$21 & 1.21$\pm$0.18\\
\hline
4C19-[\ion{C}{ii}]   & 3.5881$\pm$0.0002 & -71$\pm$12 & 256$\pm$12 & 1.62$\pm$0.06\\
\hline
\end{tabular}
\tablefoot{The reported uncertainty is the $1\sigma$ error from the fit.}
\end{table*}

\section{$^{\rm 3D}$\textsc{Barolo} fit of [\ion{C}{ii}]}\label{app:barolo}
We use $^{\rm 3D}$\textsc{Barolo} to fit the two bright [\ion{C}{ii}] companions, J0121-[\ion{C}{ii}]-B and 4C19-[\ion{C}{ii}]. In \bbarolo, a rotating galaxy disk is modeled with a set of tilted rings. Each ring is parametrized by five geometric parameters --- center coordinates ($x_{0}$, $y_{0}$), systemic velocity ($v_{\rm sys}$), position angle ($PA$, measured anti-clockwise from the North), and inclination ($i$) --- and five physical parameters --- rotation velocity ($v_{\rm rot}$), radial velocity ($v_{\rm rad}$), velocity dispersion ($\sigma_{\rm v}$), surface density ($\Sigma_{\rm gas}$), and vertical thickness ($z_{0}$). The separation between rings is set to half the width of beam major axis by default. We note that there are no constraints on the initial guesses for $PA$ and $i$. The fit is performed using values obtained from a visual inspection.

We initially perform the fit on cubes at the original resolution of 14 \kms, which leads to poorer constraints on the parameters. To mitigate this, we apply spectral smoothing to the cubes, aiming to enhance the S/N at the expense of spectral resolution. For J0121-[\ion{C}{ii}]-B, the [\ion{C}{ii}] cube is smoothed to a velocity resolution of 28 \kms. The initial values of $PA$ and $i$ are set to 193$^{\circ}$ and 60$^{\circ}$. They are allowed to be varied by $\pm15^{\circ}$. The initial values of $\sigma_{\rm v}$ is by default 8 \kms. We set \texttt{NORM=AZIM} so that the observed moment-0 map is azimuthally averaged to obtain $\Sigma_{\rm gas}$ of each ring in the model. Initial value of other parameters are guessed by \bbarolo. The center coordinates ($x_{0}$, $y_{0}$) are fixed to the initial guess. Then, the free parameters to be minimized are $v_{\rm sys}$, $PA$, $i$, $v_{\rm rot}$, and $\sigma_{\rm v}$. For 4C19-[\ion{C}{ii}], the [\ion{C}{ii}] cube is also smoothed to a velocity resolution of 28 \kms. The initial values of $PA$ and $i$ are 100$^{\circ}$ and 10$^{\circ}$. They are allowed to be varied by $\pm15^{\circ}$ and $\pm30^{\circ}$, respectively. \bbarolo\ fit is performed in two steps. First, all free parameters are fitted. Then, the geometric parameters are fixed while only $v_{\rm rot}$ and $\sigma_{\rm v}$ were fitted.

\begin{figure*}
    \centering
    \includegraphics[width=\textwidth,clip]{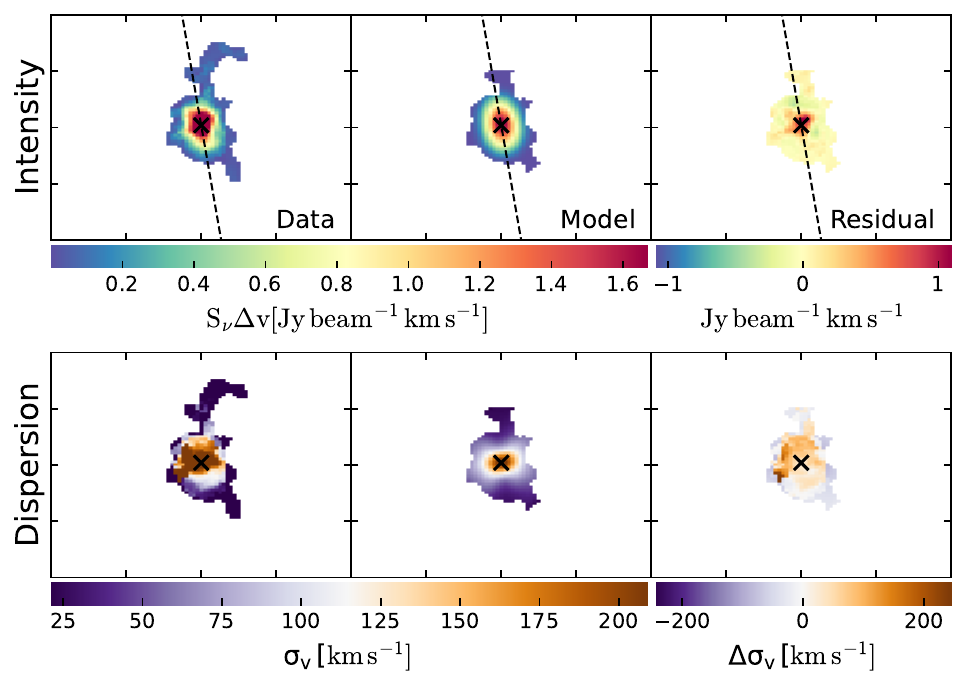}
    \caption{$^{\rm 3D}$\textsc{Barolo} moment 0 (intensity) and moment 2 (velocity dispersion) maps of J0121-[\ion{C}{ii}]-B.}
    \label{fig:barolo_m02_j0121}
\end{figure*}

\begin{figure*}
    \centering
    \includegraphics[width=\textwidth,clip]{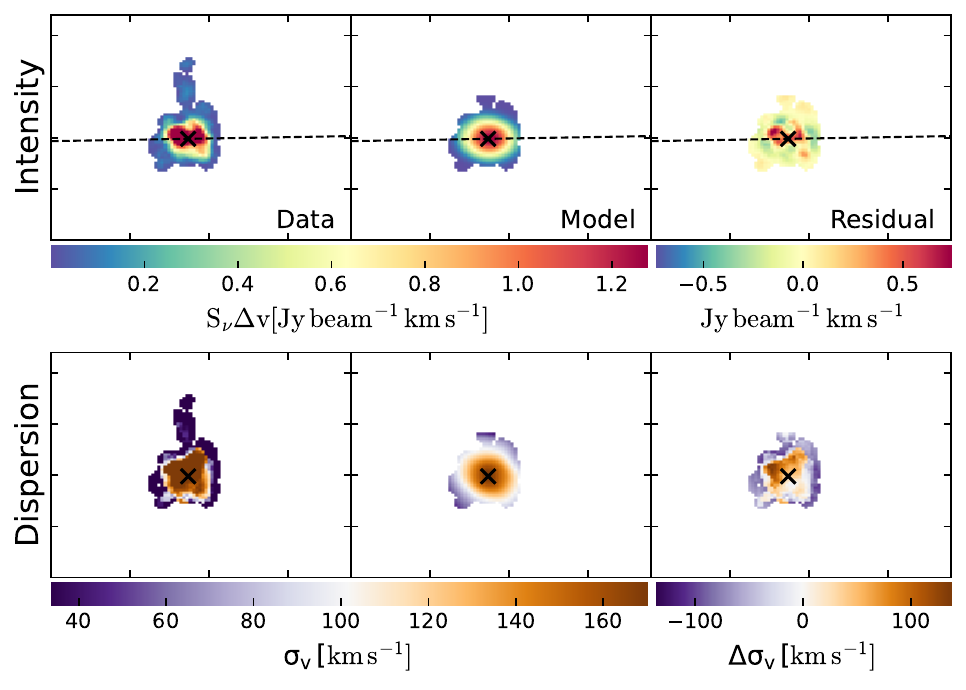}
    \caption{$^{\rm 3D}$\textsc{Barolo} moment 0 (intensity) and moment 2 (velocity dispersion) maps of 4C19-[\ion{C}{ii}].}
    \label{fig:barolo_m02_4c19}
\end{figure*}

\begin{figure}
    \centering
    \includegraphics[width=\columnwidth,clip]{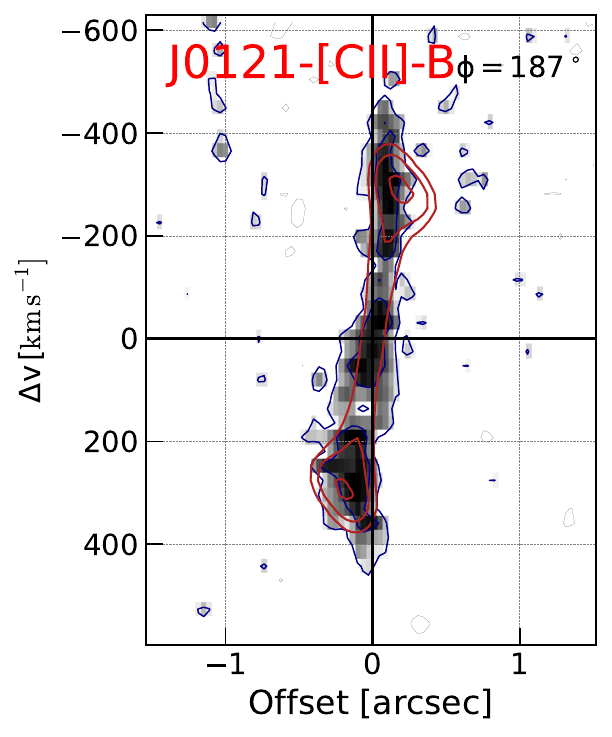}
    \caption{PV diagram of J0121-[\ion{C}{ii}]-B from $^{\rm 3D}$\textsc{Barolo} along kinematic major axis (black dashed line in Fig. \ref{fig:barolo_m1_0121}).  The gray scale maps and blue contours show the data. Red contours show the model. }
    \label{fig:barolo_pv_0121}
\end{figure}

\begin{figure}
    \centering
    \includegraphics[width=\columnwidth,clip]{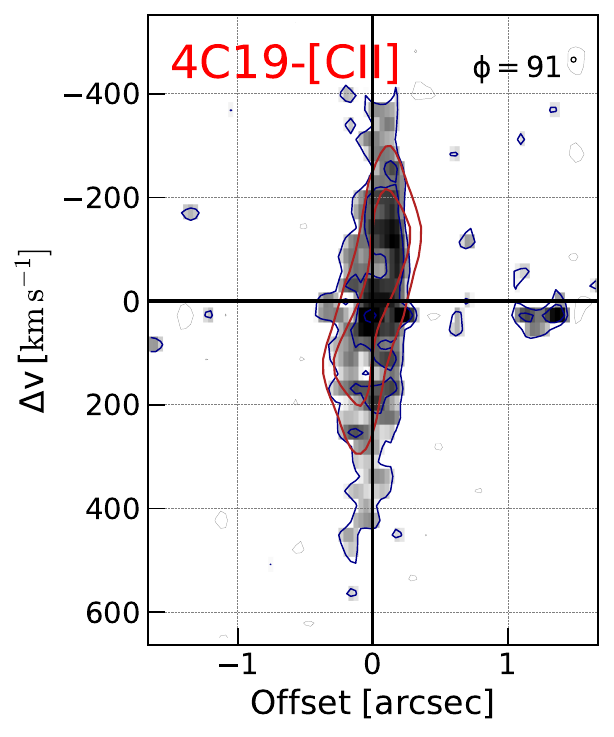}
    \caption{Similar to Fig. \ref{fig:barolo_pv_0121}, $^{\rm 3D}$\textsc{Barolo} PV diagram of 4C19-[\ion{C}{ii}].}
    \label{fig:barolo_pv_4c19}
\end{figure}

We show the fitted maps moment 0 (intensity) and moment 2 (velocity dispersion) maps in Fig. \ref{fig:barolo_m02_j0121} and \ref{fig:barolo_m02_4c19} for the two companions. In addition, we also present the position velocity (PV) diagrams from $^{\rm 3D}$\textsc{Barolo} along the kinematic major axis in Fig. \ref{fig:barolo_pv_0121} and \ref{fig:barolo_pv_4c19} for the two companions. We report that the best fits of J0121-[\ion{C}{ii}]-B $PA$ and $i$ are 187$^{\circ}$ and 61$^{\circ}$, respectively. The best fits of 4C19-[\ion{C}{ii}] $PA$ and $i$ are 93$^{\circ}$ and 30$^{\circ}$, respectively. We note that the $i$ fit of 4C19-[\ion{C}{ii}] is limited by angular resolution and should be treated with caution.

\section{Foreground detection}\label{app:foreground}

\subsection{Foreground cluster in the field of TN J0121+1320}\label{app:tnj0121_fore}
\begin{figure*}
    \centering
    \includegraphics[width=\textwidth,clip]{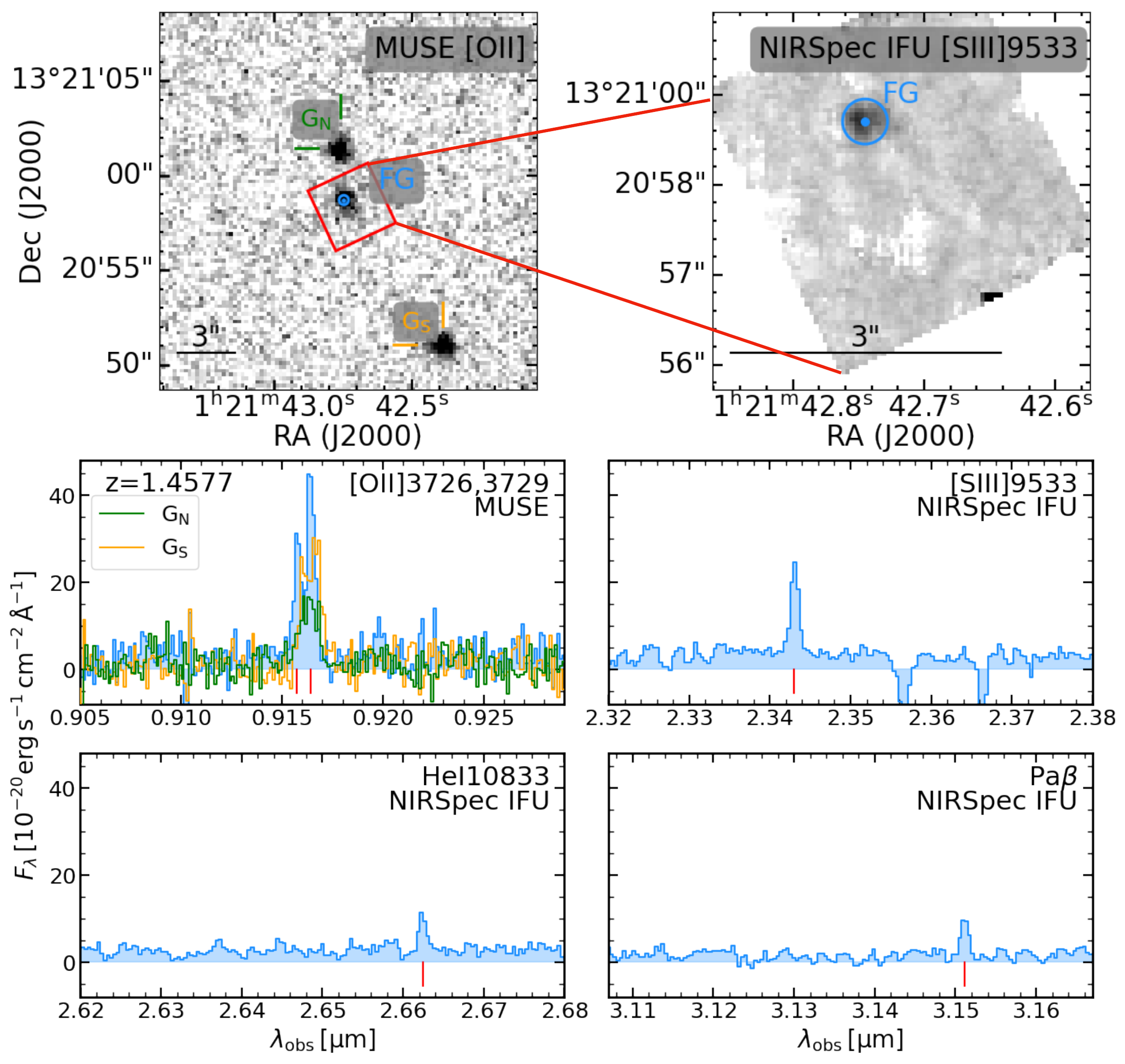}
    \caption{Narrow-band emission line images and spectra of the foreground galaxies in the field of TN J0121+1320. \textit{Upper left}: Narrow-band image of [\ion{O}{ii}]$3726,3729$ at $z=1.4577$ integrated from MUSE data cube. We mark the positions of the two  [\ion{O}{ii}] emitters outside the NIRSpec IFU FoV in green and orange, respectively. The blue circle indicates the $r=0.25\arcsec$ aperture used to extract the foreground galaxy (FG) in the NIRSpec IFU data. It is centered (blue dot) at the spatial position of the emission lines (Sect. \ref{subsec:comp_0121}) of FG, as in Fig. \ref{fig:O3_chanmap_0121}. The red box shows the FoV of the NIRSpec IFU. \textit{Upper right}: Narrow-band image of [\ion{S}{III}]$9533$ at $z=1.4577$ observed by NIRSpec IFU. The blue dot and circle are the same as in the upper left panel. \textit{Middle and lower rows}:Spectra extracted at the position of the FG in blue. We also show MUSE spectra of the two [\ion{O}{ii}] emitters in green and orange, respectively.}
    \label{fig:foreg_tnj0121}
\end{figure*}

We detect a foreground galaxy at $z=1.4577$ in the field of \object{TN J0121+1320} using the NIRSpec IFU data. The continuum emission of this foreground galaxy may contaminate the continuum of \object{TN J0121+1320} given its small spatial offset, $\sim0.1\arcsec$ (0.85$\,$kpc), from the line emission peak. This galaxy likely belongs to a cluster. Its [\ion{O}{ii}]$3726,3729$ doublet is covered by MUSE (FoV of $1\arcmin \times 1\arcmin$). We present the continuum-subtracted MUSE narrow band image integrated from $-200$ to $200\,\mathrm{km\,s^{-1}}$ around [\ion{O}{ii}]$3729$ in Fig. \ref{fig:foreg_tnj0121}. There are at least two additional [\ion{O}{ii}]$3726,3729$ emitters at similar redshift within $10\arcsec$ ($\sim84.5\,$kpc) of the one covered in the NIRSpec IFU FoV. We mark the positions of these two [\ion{O}{ii}]$3726,3729$ emitters and name them $\mathrm{G_{N}}$ and $\mathrm{G_{S}}$ for the one in the north and south, respectively. In addition to [\ion{O}{ii}]$3726,3729$, we also detect [\ion{S}{III}]$9533$, \ion{He}{I}$10833$ and Pa$\beta$ at the same redshift and position as the foreground galaxy using NIRSpec IFU. These assure the redshift of this foreground galaxy. The spectra are shown in Fig. \ref{fig:foreg_tnj0121}. The spatial position of the foreground cluster and its redshift may lead to magnification and/or distortion of the emission from TN J0121+1320 by gravitational lensing. 

\subsection{4C+03.24}\label{app:4C03_fore}
\begin{figure*}
    \centering
    \includegraphics[width=\textwidth,clip]{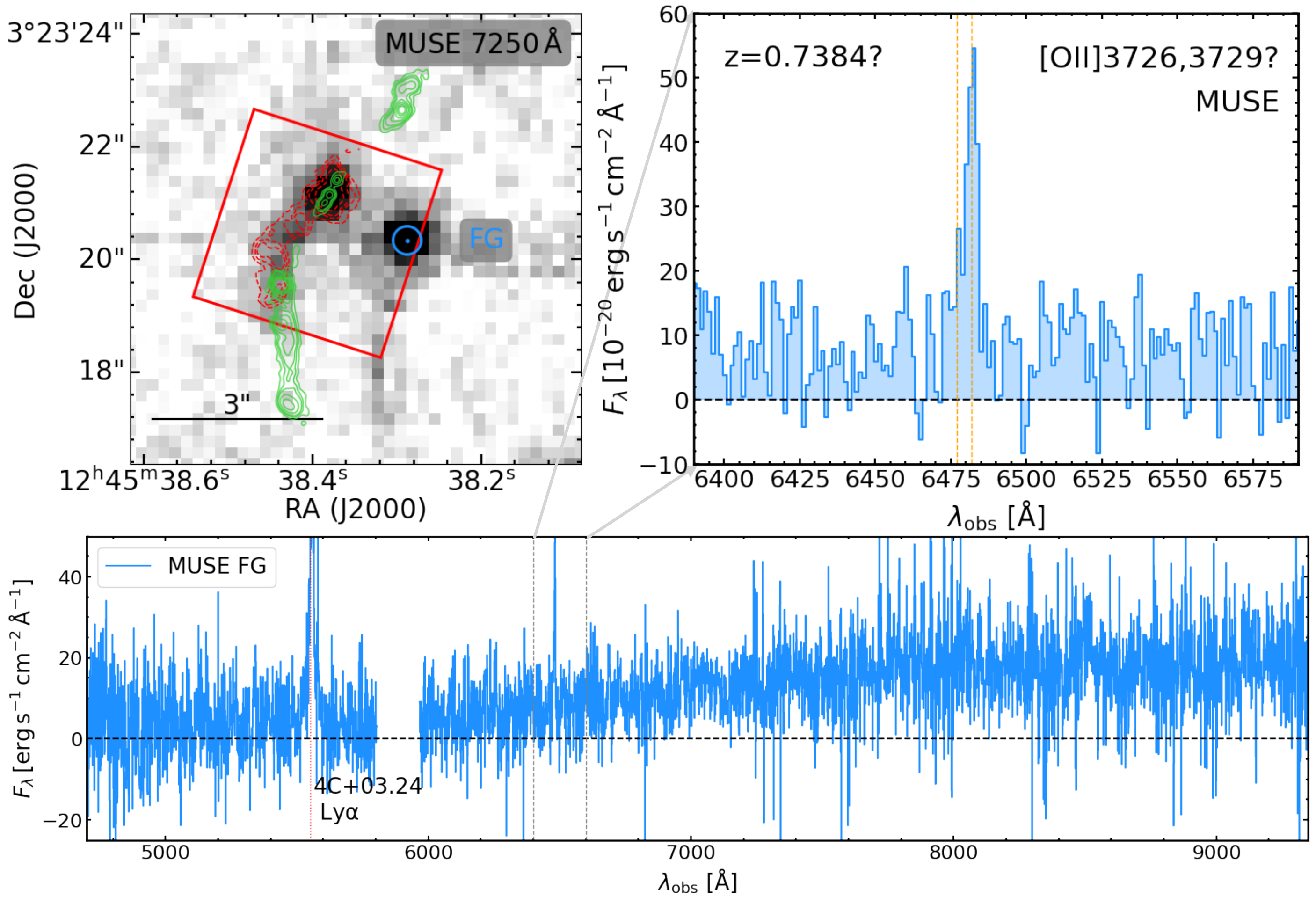}
    \caption{MUSE continuum image and spectra of foreground galaxy in the field of \object{4C+03.24}. \textit{Upper left}: Continuum image extracted from MUSE cube around $7250\,\AA$. The blue dot marks the position of the potential foreground continuum source. The blue circle shows the $r=0.25\arcsec$ aperture from which the spectrum was extracted. The red box indicates the NIRSpec IFU FoV. The green contours are same radio map as in Fig. \ref{fig:O3_chanmap_4c03}. The dashed red contours are the same NIRSpec IFU optical continuum as in Fig. \ref{fig:O3_chanmap_4c03}. \textit{Lower panel}: Full MUSE spectrum extracted at the foreground galaxy position with a focus on the continuum. The red dotted vertical line indicates $\rm Ly\alpha$ from \object{4C+03.24}. The vertical gray dashed lines mark the zoom-in region around the emission line of the foreground galaxy. \textit{Upper right}: Zoom-in view of the potential foreground [\ion{O}{II}]$3726,3729$ line.}
    \label{fig:foreg_4c03}
\end{figure*}

We report a potential foreground galaxy in the field of \object{4C+03.24}. This galaxy is covered at the edge of the NIRSpec IFU FoV. We do not find any emission lines in the NIRSpec IFU data cube and note that analysis of SINFONI data assumed it to be a star \citep[][]{Nesvadba_2017b}. At the same position in  MUSE data cube, we detect both an emission line and continuum (Fig. \ref{fig:foreg_4c03}). With only a single line detected, the redshift is uncertain. If we assume it to be redshifted [\ion{O}{ii}]$3726,3729$, the redshift is $z\sim0.7384$. The projected distance between the radio core of \object{4C+03.24} and this potential foreground object is $0.58\arcsec$, implying the possibility of gravitational lensing.

\end{appendix}
\end{document}